\documentclass[12pt]{article}

\usepackage[top=80pt,bottom=85pt,left=67pt,right=67pt]{geometry}
\usepackage{amssymb,amsmath}
\usepackage{graphicx,color}
\usepackage{float}
\usepackage{cite}
\usepackage[debug,pageanchor=false]{hyperref}
\definecolor{link}{rgb}{.8,.15,.1}
\hypersetup{colorlinks=true,linkcolor=link,citecolor=link,urlcolor=link,linktocpage}

\makeatletter
\@addtoreset{equation}{section}
\makeatother

\renewcommand{\theequation}{\thesection.\arabic{equation}}

\newcommand{\beq}{\begin{equation}}
\newcommand{\eeq}{\end{equation}}
\newcommand{\bea}{\begin{eqnarray}}
\newcommand{\eea}{\end{eqnarray}}

\begin{document}
\begin{titlepage}

\begin{center}

\vskip .5in 
\noindent

{\Large \bf{Holographic Description of M-branes via AdS$_2$}}

\bigskip\medskip

Giuseppe Dibitetto\footnote{dibitettogiuseppe@uniovi.es}, Yolanda Lozano\footnote{ylozano@uniovi.es},  Nicol\`o Petri\footnote{petrinicolo@uniovi.es}, Anayeli Ramirez\footnote{ramirezanayeli.uo@uniovi.es} \\

\bigskip\medskip
{\small 

Department of Physics, University of Oviedo,
Avda. Federico Garcia Lorca s/n, 33007 Oviedo, Spain}

\vskip 1.5cm 
\vskip .9cm 
     	{\bf Abstract }

\vskip .1in
\end{center}

\noindent
We study $\textrm{AdS}_2\times S^4 \times S^2 \times \Sigma_2$ solutions in type IIB string theory arising from D1 -- D3 -- NS5 brane intersections. These backgrounds enjoy sixteen supercharges and the corresponding internal geometry is non-compact due to the specific form of the warping w.r.t. the Riemann surface $\Sigma_2$. Even though a direct computation of the holographic free energy of the would-be dual CFT$_1$ yields a divergent behaviour, it reveals the typical $N^3$ scaling of a 6d theory upon introducing a hard cut-off. We claim that such geometries may be interpreted as the gravity duals of M(atrix) models describing an IR  phase of the $(2,0)$ theory of M5 branes, in presence of momentum and NUT charges.
We discuss parallel $\textrm{AdS}_2$ geometries describing longitudinal M2 branes in the UV,
where the counting of the number of degrees of freedom correctly reproduces the expected $N^{3/2}$ behaviour  of the dual field theory. These geometries provide explicit examples where  deconstructed extra dimensions yield well-defined UV descriptions in terms of higher-dimensional CFTs.

\noindent
 
\vfill
\eject

\end{titlepage}

\setcounter{footnote}{0}

\tableofcontents

\setcounter{footnote}{0}
\renewcommand{\theequation}{{\rm\thesection.\arabic{equation}}}

\section{Introduction}

Ever since the advent of the AdS/CFT correspondence in the context of type IIB string theory \cite{Maldacena:1997re}, it has become increasingly important to extend our encyclopedic knowledge of supergravity backgrounds involving AdS factors. The opportunity of relating classical computations in AdS to field theory results in the regime of strong coupling \cite{Witten:1998qj} has enormously increased our understanding of non-perturbative effects in quantum field theory. In particular, one of the important lessons that we learned from this is the existence of non-Lagrangian phases of quantum fields in which any perturbative treatment is bound to fail.

From the gravity side, the R-symmetry of a supersymmetric conformal field theory (CFT) turns out to be geometrically realised as an isometry of the internal manifold. Furthermore, in the holographic limit the amount of dynamical degrees of freedom within the CFT is \;proportional to the effective lower-dimensional Newton constant in AdS \cite{Brown:1986nw}, which is in turn related to the volume of the internal manifold. 
Hence classifying the range of different CFTs compatible with a given amount of supersymmetry translates into the scan of all possible geometrical structures and holonomies of the corresponding internal manifolds.

As a consequence, the higher the dimensionality of the internal space, the richer the structure of all possible geometries and topologies thereof becomes. This makes it increasingly challenging to exhaustively classify AdS vacua with lower and lower dimensionality. In\;particular, starting off by the highest dimension possible \cite{Nahm:1977tg} (i.e. seven), an exhaustive classification of supersymmetric AdS$_d$ solutions has been achieved for every $d$ up to four (see e.g. refs \cite{Grana:2004bg,Lust:2004ig,Gauntlett:2004zh,Gauntlett:2004hs,Bovy:2005qq,Grana:2006kf,Kounnas:2007dd,Kim:2007hv,Tomasiello:2007eq,Koerber:2008rx,Caviezel:2008ik,Apruzzi:2013yva,Apruzzi:2014qva,Kim:2015hya,DHoker:2016ujz,DHoker:2016ysh,DHoker:2017mds,DHoker:2017zwj,Passias:2017yke,Passias:2018zlm}).
When going further down to $d=3$ and $2$, while it is arguably hard to exhaust all possibilities, partial attempts of classifications restrict the analysis to specific classes of geometries \cite{Argurio:2000tg,Kim:2005ez,Gauntlett:2006ns,Gauntlett:2006af,DHoker:2007mci,Donos:2008hd,Corbino:2017tfl,Corbino:2018fwb,Dibitetto:2017tve,Kelekci:2016uqv,Dibitetto:2017klx,Couzens:2017way,Eberhardt:2017uup,Couzens:2017nnr,Gauntlett:2018dpc,Dibitetto:2018gbk,Dibitetto:2018ftj,Dibitetto:2018iar,Dibitetto:2018gtk,Couzens:2018wnk,Gauntlett:2019roi,Macpherson:2018mif,Hong:2019wyi,Lozano:2019emq,Lozano:2019jza,Couzens:2019iog}. 

Focussing our attention on the case of AdS$_2$, besides presenting technical challenges,\;\; AdS$_2$/CFT$_1$ has been argued to pose far more urgent conceptual puzzles  \cite{Maldacena:1998uz,Denef:2007yt,Maldacena:2016hyu,Maldacena:2016upp}. These issues mainly originate from the crucial fact that the boundary of AdS$_2$ spacetime is non-connected, its topology being the one of two disjoint lines. This feature was identified  in \cite{Harlow:2018tqv} as the origin of a non-factorisability of the quantum gravity partition function in AdS$_2$, hence causing an unsolvable holographic mismatch. 

A possible way out of this conundrum could be to avoid insisting in retaining an AdS$_2$/CFT$_1$ correspondence which is supposed to be valid all the way to the UV, and investigate instead the possibility that its UV completion might resolve it into higher dimensions.
In this work we take this perspective and investigate it in detail within a specific string theory set-up, where the impossibility to decouple the 2d gravitational degrees of freedom from the bulk stems from the non-compactness of the corresponding internal manifold.  

We will start from an example of solution in type IIB supergravity belonging to the class studied in \cite{DHoker:2007mci}, whose 10d geometry is given by $\mathrm{AdS}_2\times S^4\times S^2$ warped over a compact Riemann surface $\Sigma_2$, and firstly show how it can be obtained as the near horizon limit of a semilocalised D1 -- NS5 -- D3 intersection. The geometry of the corresponding AdS$_2$ vacuum is non-compact, and hence a naive calculation of the holographic free energy of the would-be dual SCFT$_1$ yields a divergent result. 
In this special set-up we will still be able to argue for the existence of a superconformal quantum mechanics (SCQM) describing this system in the IR regime, while we will show how this pathological behaviour of the free energy is cured in the UV by the emergence of deconstructed extra dimensions yielding a UV description in terms of a higher-dimensional CFT.

In order to support this claim, we will perform a single (Abelian) T-duality to obtain \;another AdS$_2$ geometry, this time in type IIA, whose underlying brane set-up will now involve D0, as well as KK5 and D4-branes. However, once in IIA it is possible to appeal to an M-theory description to have the divergence cured. Indeed what we will see is that the corresponding 11d geometry is nothing but $\mathrm{AdS}_7/\left(\mathbb{Z}_{k}\times\mathbb{Z}_{k'}\right)\times S^4$, describing a stack of M5-branes carrying momentum and NUT charges. The introduction of an explicit cut-off regulator in the holographic free energy calculated earlier in IIA (as well as in IIB) will identify an IR phase describing a D-particle gas, whose free energy will scale as $k^2$, where $k$ is the number of D0-branes. On the other hand, in a UV regime we will rather recover the typical $N^3$ scaling of a 6d theory describing a stack of $N$ M5-branes.

Subsequently, in order to further test our novel insight, we will repeat a similar analysis for another non-compact AdS$_2$ solution in type IIA supergravity, arising from the near horizon limit of a D0 -- F1 intersection \cite{Cvetic:2000cj}. The corresponding geometry here is given by $\mathrm{AdS}_2 \times S^7$ warped over an interval $I_\alpha$.
The 11d uplift of this background is now given by $\mathrm{AdS}_4/\mathbb{Z}_k\times S^7$, with the non-compactness of the 10d background hidden within a 4d AdS geometry. The same  logic turns out to go through again, with two different regimes emerging. 
The IR phase looks very similar to the one previously discussed, which should not come as a surprise since it still describes a D0-brane QM, while its UV completion is now given in terms of longitudinal M2-branes.

In conclusion, we will provide here suitable candidate realisations of an AdS$_2$/CFT$_1$ correspondence, within explicit controlled string theory set-ups. The important feature that precisely allows such constructions to hold is non-compactness. This of course also sets its own limitations in that it requires inserting a cut-off. On the field theory side, it represents the energy scale up to which we expect our 1d description to be valid. On the gravity side, it corresponds to probing a regime where the internal geometry is artificially made compact. Note that this exactly coincides with inserting back gravitational degrees of freedom in $\mathrm{AdS}_2$, which are otherwise absent in a non-compact situation.

Finally, we will briefly discuss another D1 -- NS5 -- D3 solution in type IIB in the class in \cite{DHoker:2007mci} with a non-compact Riemann surface. This solution is interesting in that it suggests that wide classes of 1d CFTs should exist emerging as IR fixed points of quantum mechanics described by linear quivers.
 The detailed study of these CFTs and their (possibly higher-dimensional) UV completion is left however for future work. 

The organisation of the paper is as follows. In section \ref{AdS2IIB} we construct the $\mathrm{AdS}_2$ solution from D1 -- NS5 -- D3 branes that is the main focus of this paper. We discuss its type IIA realisation as well as its uplift to M-theory. In section \ref{field-theory} we describe its dual interpretation as a M(atrix) model describing the IR phase of the 6d CFT describing M5-branes with momentum and NUT charges. In section \ref{D0-D1} we discuss the 11d uplift of the $\mathrm{AdS}_2\times S^7$ solution arising from the near horizon of D0 -- F1, and provide an interpretation as a D0 quantum mechanics completed in the UV by longitudinal M2-branes. In section \ref{NATD} we discuss a type IIB solution arising from a D1 -- NS5 -- D3 intersection dual to a 1d CFT described by a linear quiver. In section \ref{conclusions} we summarise our results and discuss future directions. In Appendix \ref{analytic} we show that the $\mathrm{AdS}_2$ solution discussed in section 2 is related to a IIB $\mathrm{AdS}_4\times S^2\times S^2\times \Sigma_2$ solution in the class of \cite{DHoker:2007zhm,DHoker:2007hhe} through an analytic continuation prescription.

\section{$\mathrm{AdS}_2$ Solutions from D1 -- NS5 -- D3 Branes} \label{AdS2IIB}
   
In this section we show that a given class of AdS$_2$ backgrounds arising from D1 -- NS5 -- D3 brane intersections provides a holographic low energy description of M(atrix) models for M5 branes. We start  discussing the brane setup in type IIB, together with the physical properties of the corresponding AdS solutions, including their holographic free energy. Due to non-compactness, this will yield a divergent result. In order to understand the physical meaning of this singular behaviour, we will then move to a type IIA picture obtained by performing a single T-duality. The brane system will be now given by intersecting D0, KK5 and D4-branes, while the related near horizon geometry will still be given by the warped product of AdS$_2$ and a non-compact 8-manifold. This non-compactness issue is finally resolved by lifting this solution to eleven dimensions, where it is identified as the near horizon geometry of an M5 brane stack with extra momentum and NUT charges. Our results in this section confirm previous connections in the literature between M(atrix) models and the AdS/CFT correspondence (see for instance \cite{Hyun:1998bi,Hyun:1998iq,Awata:1998qy,Berkooz:1999iz}).

   \subsection{The Type IIB Picture} \label{TypeIIB_Picture}
We start from the D1 -- NS5 -- D3 brane intersection described in Table \ref{D1-NS5-D3}. 
It can be shown to preserve eight real supercharges, as well as $\mathrm{SO}(5)\times \mathrm{SO}(3)$ bosonic symmetry. The corresponding field theory description is an $\mathcal{N}=8$ supersymmetric quantum mechanics whose supercharges transform as spinors of the above R-symmetry group, i.e. in the $(\textbf{4},\textbf{2})$. In particular, 
the $\mathrm{SO}(5)$ factor emerges in the Coulomb branch of the 1d theory, while the extra $\mathrm{SO}(3)$ becomes manifest in the Higgs branch.
  \begin{table}[ht]
	\begin{center}
		\begin{tabular}{| l | c | c | c | c| c | c| c | c| c | c |}
			\hline		    
			& 0 & 1 & 2 & 3 & 4 & 5 & 6 & 7 & 8 & 9 \\ \hline
			D1 & $\times$ &  & &  &  &  & $\times$  &   &   &   \\ \hline
			NS5 & $\times$ & $\times$ &$\times$  &$\times$  & $\times$ &$\times$   &  &  &  &   \\ \hline
			D3 & $\times$ &  &  &  &  &  &   & $\times$  &$\times$   &$\times$   \\ \hline
		\end{tabular} 
	\end{center}
	\caption{The $\frac14$-BPS intersection involving D1, D3 and NS5 branes. A supersymmetric quantum mechanics lives in the common $x^0$ direction. The corresponding $\mathrm{SO}(5)_R\times \mathrm{SO}(3)_R$ R-symmetry is geometrically realised as rotations in the $(x^1, \dots, x^5)$ and $(x^7,x^8,x^9)$ coordinates, respectively.   An $\mathrm{AdS}_2$ vacuum will be obtained by taking the near horizon limit of this system and a superconformal quantum mechanics will arise in the IR limit of the QM.}   
	\label{D1-NS5-D3}	
\end{table} 

The corresponding type IIB supergravity background reads
\begin{equation}
   \begin{split}
    d s_{10}^2&=-H_{\mathrm{D}1}^{-1/2}\,H_{\mathrm{D}3}^{-1/2}\,d t^2+H_{\mathrm{D}1}^{1/2}\,H_{\mathrm{D}3}^{1/2}\,\left(d \rho^2+\rho^2\,d s^2_{S^4} \right)+H_{\mathrm{D}1}^{-1/2}\,H_{\mathrm{D}3}^{1/2}\,H_{\mathrm{NS}5}\,d y^2\,+\\
   &\phantom{=}\,+\,H_{\mathrm{D}1}^{1/2}\,H_{\mathrm{D}3}^{-1/2}\,H_{\mathrm{NS}5}\,\left(d r^2+r^2\,d s^2_{S^2} \right) \ ,\\
   e^{\Phi}&=H_{\mathrm{D}1}^{1/2}\,H_{\mathrm{NS}5}^{1/2}\ ,\qquad \ \ \, B_{(6)}=\frac{H_{\mathrm{D}3}}{H_{\mathrm{NS}5}}\, dt \wedge \mathrm{vol}_{\left(\mathbb{R}^5\right)}\ ,\\
   C_{(2)}&=\frac{1}{H_{\mathrm{D}1}}\, dt \wedge dy\ ,\qquad C_{(4)}=\frac{H_{\mathrm{NS}5}}{H_{\mathrm{D}3}}\, dt \wedge \mathrm{vol}_{\left(\mathbb{R}^3\right)}\ ,
   \label{D1D3NS5_metric}
   \end{split}
  \end{equation}
where we denoted by $\rho$ the radial coordinate of $\mathbb{R}^5$ parameterised by $(x^1,\dots,x^5)$ and $r$ the one of $\mathbb{R}^3$ parameterised by $(x^7,x^8,x^9)$. As a consequence, $H_{\mathrm{D}1}=H_{\mathrm{D}1}(\rho,r)$, $H_{\mathrm{D}3}=H_{\mathrm{D}3}(\rho)$ and $H_{\mathrm{NS}5}=H_{\mathrm{NS}5}(r)$ are suitable functions. The equations of motion of type IIB supergravity plus the Bianchi identities are satisfied by the following explicit form of the aforementioned functions
\begin{equation}
H_{\mathrm{D}1}=1+Q_{\mathrm{D}1}^{-2}\left(\frac{\pi Q_{\mathrm{D}3}}{\rho}\,+Q_{\mathrm{NS}5}\,r\right)\ ,\quad H_{\mathrm{D}3}=\frac{\pi Q_{\mathrm{D}3}}{\rho^3}
\quad \text{and}\quad H_{\mathrm{NS}5}=\frac{Q_{\mathrm{NS}5}}{r}\ ,
\label{D1D3NS5_solution}
\end{equation}
where the integration constants appearing above are interpreted as quantised brane charges.
By taking a look at the form of $H_{\mathrm{D}1}$, we immediately see that the near horizon limit corresponds to the following regime
\begin{equation}
 \rho\,\ll\, Q_{\mathrm{D}3} \qquad \text{while}\qquad r\,\gg\,Q_{\mathrm{NS}5}^{-1}\ ,
\end{equation}
in such a way that both terms in $H_{\mathrm{D}1}$ compete.
In order to better understand this limit, it is useful to introduce the following change of coordinates
\begin{equation}
 \rho=Q_{\mathrm{D}3}\,\zeta\,\sin^{2}\alpha \qquad \text{and}\qquad r=Q_{\mathrm{NS}5}^{-1}\,\zeta^{-1}\,\cos^{2}\alpha\ ,\label{NH_coordinates}
\end{equation}
where $\alpha$ ranges from zero to $\frac{\pi}{2}$, while $\zeta$ goes from zero to $\infty$. In terms of these new coordinates the near horizon limit is achieved by taking $\zeta\,\rightarrow 0$.
This procedure yields the following result 
\begin{eqnarray}
ds_{10}^2&=&\frac{\ell^2}{\sin^3{\alpha}}\left(ds^2_{\mathrm{AdS}_2}+4d\alpha^2+\sin^2{\alpha}\,ds^2_{S^4}+\cos^2{\alpha}\,ds^2_{S^2}+R_0^2\, \frac{\sin^6{\alpha}}{\cos^2{\alpha}}\,dy^2\right)\ ,\label{NH_D1D3NS5}\\
e^\Phi&=&\frac{Q_{\mathrm{NS}5}}{Q_{\mathrm{D}1}}\cos^{-1}{\alpha}\ ,\\
H_{(3)}&=&-Q_{\mathrm{NS}5}\,dy\wedge {\rm vol}_{\left(S^2\right)}\ ,\\
F_{(3)}&=&-{Q_{\mathrm{D}1}}\,{\rm vol}_{\left(\mathrm{AdS}_2\right)}\wedge dy\ ,\\
F_{(5)}&=&-3\pi Q_{\mathrm{D}3}\,{\rm vol}_{\left(S^4\right)}\wedge dy+6\pi\ell^2\,\frac{ Q_{\mathrm{D}3}}{Q_{\mathrm{NS}5}}\frac{\cos^3(\alpha)}{\sin^7(\alpha)}\,d\alpha\wedge {\rm vol}_{\left(\mathrm{AdS}_2\right)}\wedge {\rm vol}_{\left(S^2\right)}\ , \label{D1D3NS5_horizon}
\end{eqnarray}
where 
\begin{equation}
\ell^2=\frac{\pi Q_{\mathrm{D}3}}{Q_{\mathrm{D}1}}\ , \qquad R_0=\frac{Q_{\mathrm{NS}5}\,Q_{\mathrm{D}1}}{\pi Q_{\mathrm{D}3}}\ ,
\end{equation}
$\alpha\in \left[0,\frac{\pi}{2}\right]$ and $y\in [0,\pi]$\footnote{This choice of parameterisation of the $S_y^1$ reproduces in the type IIA picture the right periodicity of the Hopf fiber coordinate.}, and the D1, D3 and NS5 brane charges appearing above are integers once measured in string units. This metric represents a foliation of $\mathrm{AdS}_2\times S^4 \times S^2 \times S^1_y$ with warping over $I_\alpha$, and belongs to the class of $\mathrm{AdS}_2\times S^4 \times S^2 \times \Sigma_2$ solutions studied in \cite{DHoker:2007mci}, where the Riemann surface $\Sigma_2$ is given by $I_\alpha \times S^1_y$, as we show below.
It is worth mentioning that the above background is made non-compact by the specific form of the warping. As a consequence the 7-form flux,
\begin{equation}
F_{(7)}=2\ell^6 \frac{Q_{\mathrm{D}1}}{Q_{\mathrm{NS}5}}\frac{\cos^3\alpha}{\sin^5\alpha}\,d\alpha\wedge {\rm vol}_{\left(S^2\right)}\wedge {\rm vol}_{\left(S^4\right)}\ ,
\end{equation}
yields an infinite result for the magnetic D1-brane charge, upon integration along the warping coordinate $\alpha$. We will give an interpretation to this divergence when we discuss below the holographic free energy.

The isometries of the solution are given by $\mathrm{SO}(2,1)\times \mathrm{SO}(5)\times \mathrm{SO}(3)$, which should match the spacetime and R-symmetries of a dual superconformal quantum mechanics with 8 supercharges. The superconformal groups with 8 supercharges containing $\mathrm{SO}(2,1)$ as a bosonic subgroup were classified in \cite{Frappat:1996pb}. We list them in Table~\ref{groups}, together with the corresponding R-symmetries.
 \begin{table}[ht]
	\begin{center}
		\begin{tabular}{ | c | c |}
			\hline		    
			 Supergroup & R-symmetry  \\ \hline
			$\mathrm{OSp}(8|2)$ & $\mathrm{SO}(8)$   \\ \hline
			$\mathrm{SU}(1,1|4)$ & $\mathrm{SU}(4)\times \mathrm{U}(1)$   \\ \hline
			$\mathrm{OSp}(4^*|4)$ & $\mathrm{SU}(2)\times \mathrm{SO}(5)$    \\ \hline
			$\mathrm{F}(4)$ & $\mathrm{SO}(7)$ \\ \hline
		\end{tabular} 
	\end{center}
	\caption{Superconformal algebras with $\mathcal{N}=8$ supersymmetry containing $\mathrm{SO}(1,2)\times G_R \, = \,\mathrm{ISO}(\mathrm{AdS}_2) \times G_R$. The various R-symmetry groups $G_R$ are realised geometrically in the corresponding supergravity duals. Therefore, the relevant one in our case is $\mathrm{OSp}(4^*|4)$.}
	\label{groups}	
\end{table} 
 We see that the $\mathrm{SO}(5)\times \mathrm{SO}(3)$ part of the isometry group matches the R-symmetry of an $\mathrm{OSp}(4^*|4)$ supergroup. The type IIB supergravity solutions realising the $\mathrm{OSp}(4^*|4)$ supergroup turn out to be precisely the $\mathrm{AdS}_2\times S^4\times S^2\times \Sigma_2$ geometries classified in \cite{DHoker:2007mci}, to which our solution belongs. There, each AdS$_2$ solution is specified by the choice of two harmonic functions on $\Sigma_2$. 
The specific choice of harmonic functions underlying our background is given by 
 \begin{equation}
 h_1=\frac{\ell^2}{2} \frac{Q_{\mathrm{D}1}}{Q_{\mathrm{NS}5}}\cot^2{\alpha}\, \qquad h_2=\frac{\ell^2}{2} \sin^{-2}{\alpha},
 \end{equation}
while for the fluxes we find 
 \begin{equation}
 h_1^D=-\frac12 Q_{\mathrm{D}1} y\, , \qquad h_2^D=-\frac12 Q_{\mathrm{NS}5} y\, .
 \end{equation}
 Note that due to the non-compactness of our solution some of the regularity conditions imposed in \cite{DHoker:2007mci} are however not satisfied. 

 Interestingly, the $\mathrm{OSp}(4^*|4)$ supergroup allows for a different choice of real section with $\mathrm{SO}(3,2)\times \mathrm{SO}(3)\times \mathrm{SO}(3)$ bosonic subgroup. This is the group of isometries of the $\mathrm{AdS}_4\times S^2\times S^2\times \Sigma_2$ solutions constructed in \cite{DHoker:2007zhm,DHoker:2007hhe}. This strongly suggests that the $\mathrm{AdS}_2\times S^4 \times S^2$ solution should be related to a solution in this class upon analytical continuation. This solution and the corresponding harmonic functions are given in the Appendix.

\vspace{2mm}
\noindent \textbf{Brane singularities:} The metric given in \eqref{NH_D1D3NS5} exhibits singularities at both boundaries of the range of $\alpha$. As $\alpha\rightarrow 0$, the dilaton asymptotes to a constant and hence one expects a D3 brane singularity. Indeed, after introducing $\beta=\alpha^{-2}$, the metric reads
\begin{equation}
ds_{10}^2\sim\,\ell^2\left[\beta^{3/2}\left(ds^2_{\mathrm{AdS}_2}+ds^2_{S^2}\right)\,+\,\beta^{-3/2}\left(d\beta^2\,+\,\beta^2\,ds^2_{S^4}\,+\,R_0^2\,dy^2\right)\right] \ ,\label{D3_singularity}\\ 
\end{equation}
which correctly reproduces the metric of a D3 brane with worldvolume $\mathrm{AdS}_2\times S^2$, localised in the origin of $\mathbb{R}^5$ and smeared over the $y$ circle. On the other hand, as $\alpha\rightarrow \frac{\pi}{2}$, the metric takes the form
\begin{equation}
ds_{10}^2\sim\,\ell^2\left[\left(ds^2_{\mathrm{AdS}_2}+ds^2_{S^4}\right)\,+\,\beta^{-1}\left(d\beta^2\,+\,\beta^2\,ds^2_{S^2}\,+\,R_0^2\,dy^2\right)\right] \ ,\label{NS5_singularity}\\ 
\end{equation}
where $\beta=\left(\frac{\pi}{2}-\alpha\right)^2$, while the dilaton behaves as $e^\Phi\sim\,e^\Phi_0\,\beta^{-1/2}$. This is the typical form of an NS5 brane singularity with a worldvolume given by $\mathrm{AdS}_2\times S^4$, localised at the origin of $\mathbb{R}^3$ and smeared over the $y$ circle.

\begin{figure}
\centering
\includegraphics[scale=0.55]{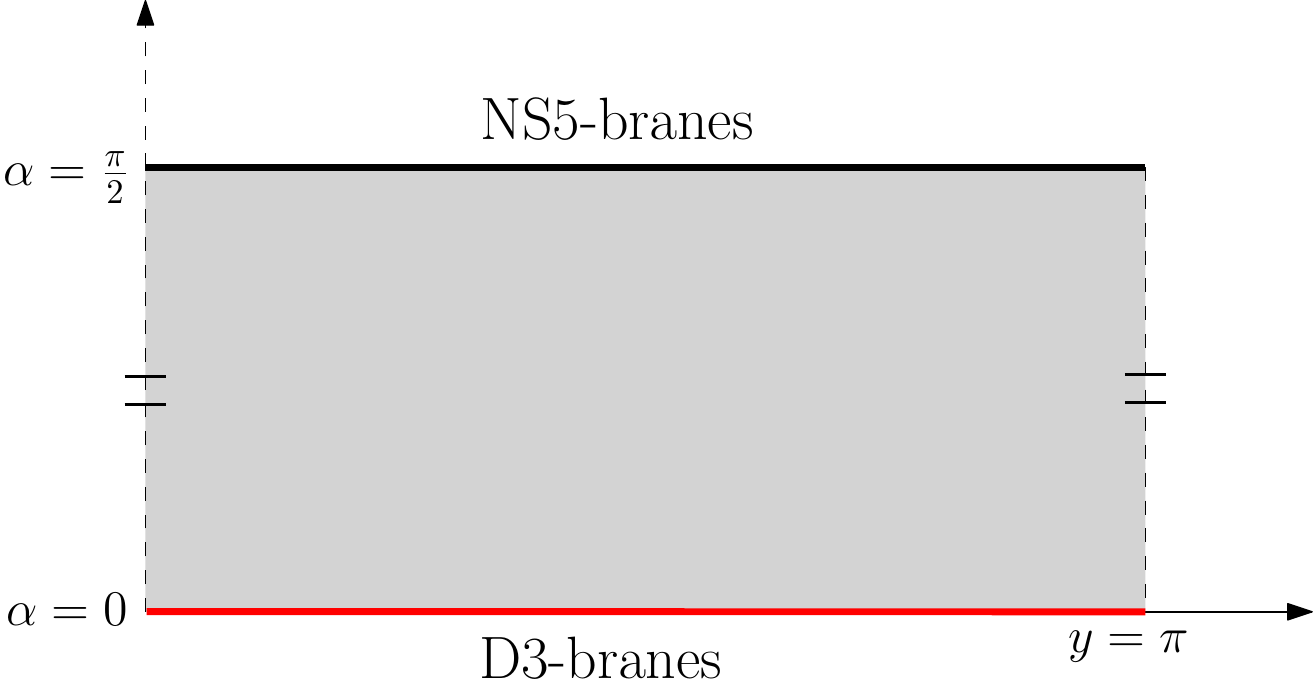}
\caption{Annulus associated to the solution.}
\label{Riemann1}
\end{figure}

The 2d Riemann surface associated to the solution is the annulus depicted in Figure \ref{Riemann1}. D3-branes are smeared over the lower boundary at $\alpha=0$ and NS5-branes are smeared along the upper boundary at $\alpha=\frac{\pi}{2}$. The annulus topology follows from the periodicity under $y\rightarrow y + \pi$.

\vspace{2mm}
\noindent \textbf{Holographic free energy:} In order to evaluate the effective number of degrees of freedom of the dual 1d theory, we follow the standard prescription that relates it to the inverse effective Newton constant in the 2d gravity dual. Following the prescription in \cite{Klebanov:2007ws,Macpherson:2014eza,Bea:2015fja} we have that, 
for a generic dilaton and  background of the form,
\begin{equation}
ds_{10}^2= a(\zeta,\vec{\theta})\left(dx_{1,d}^2 + b(\zeta)d\zeta^2\right) + g_{ij}(\zeta,\vec{\theta}) d\theta^id\theta^j,\;\;\;\; \Phi=\Phi(\zeta,\vec{\theta}),
\end{equation}
the free energy is computed from the auxiliary quantity 
\begin{equation}
\hat{H}= \left(\int d\vec{\theta} \sqrt{e^{-4\Phi} \det[g_{ij}] a(\zeta,\vec{\theta})^d } \right)^2,\label{Hhat}
\end{equation}
as 
\begin{equation}
{\cal F}_{\mathrm{hol}}= 3\times \frac{ d^d}{ G_{\mathrm{N}}} \frac{b(\zeta)^{d/2} (\hat{H})^\frac{2d+1}{2}  }{(\hat{H}')^d} ,\label{centralx}
\end{equation}
where $G_{\mathrm{N}}$ is the Newton's constant in ten dimensions, $G_{\mathrm{N}}=8\pi^6$.
For the case at hand 
\begin{equation}
d=0\, , \qquad a(\zeta,\vec{\theta})=\frac{\ell^2}{\sin^3{\alpha}}\zeta^2\, , \qquad b(\zeta)=\frac{1}{\zeta^4}
\end{equation}
and we obtain
\begin{equation}
\sqrt{\hat{H}}=\frac{16}{3}\pi^7 \frac{Q_{\mathrm{D}3}^3}{Q_{\mathrm{D}1}\,Q_{\mathrm{NS}5}}\int_0^{\pi/2}d\alpha \,\frac{\cos^3{\alpha}}{\sin^5{\alpha}}
\end{equation}
The integral in $\alpha$ diverges close to $\alpha=0$, as a reflection of the non-compactness of the internal space. Indeed the divergence is exactly the same as that of the magnetic D1-brane charge that we mentioned above. Regularising it with a hard cut-off $\epsilon$ we find
\begin{equation}
{\cal F}_{\mathrm{hol}}=\frac{\pi}{2}\frac{Q_{\mathrm{D}3}^3}{Q_{\mathrm{D}1}\,Q_{\mathrm{NS}5}}\cot^4{\epsilon}\ .
\label{FHol_IIB}
\end{equation}
Leaving aside the singularity, the behaviour of the free energy is suggestive of a 6d CFT associated to $Q_{\mathrm{D}3}$ M5 branes seated at a $\mathbb{Z}_{Q_{\mathrm{D}1}}\times \mathbb{Z}_{Q_{\mathrm{NS}5}}$ orbifold. We show in the next sections that it is indeed possible to give such an interpretation to  the superconformal quantum mechanics dual to the solution.

\subsection{T-dual to Type IIA}\label{ATD}

The previous solution is equivalent upon Abelian T-duality on the $y$ direction to the following solution in type IIA:
\begin{eqnarray}
&&ds^2_{10}=\frac{\ell^2}{\sin^3{\alpha}}\left(ds^2_{\mathrm{AdS}_2}+4\,d\alpha^2+\sin^2{\alpha}\,ds^2_{S^4}+4\,\cos^2{\alpha}\,ds^2_{S^3/\mathbb{Z}_{k^\prime}}\right) \label{IIAmetric} \\
&&e^\Phi=\frac{\ell}{Q_{\mathrm{D}0}}\sin^{-3/2}{\alpha}\\
&&F_{(2)}=-Q_{\mathrm{D}0}{\rm vol}_{\left(\mathrm{AdS}_2\right)}\\
&&F_{(4)}=-3\pi Q_{\mathrm{D}4}{\rm vol}_{\left(S^4\right)}, \label{IIAF4}
\end{eqnarray}
where, upon T-duality, $Q_{\mathrm{D}1}=Q_{\mathrm{D}0}$, $Q_{\mathrm{D}3}=Q_{\mathrm{D}4}$ and $Q_{\mathrm{NS}5}=Q_{\mathrm{KK}5}=k^\prime$;
$\ell^2=\frac{\pi Q_{\mathrm{D}4}}{Q_{\mathrm{D}0}}$ and the 3-sphere appearing in \eqref{IIAmetric} is locally written as a Hopf fibration of a $S^2$ on $S^1_y$,
\begin{equation}
\label{S3Z}
ds^2_{S^3/\mathbb{Z}_{k^\prime}}=\frac14\left[\left(\frac{dy}{k^\prime}+\omega\right)^2+ds^2_{S^2}\right]\qquad \text{with} \qquad d\omega={\rm vol}_{\left(S^2\right)}\,.
\end{equation}
This solution can be consistently obtained as the near-horizon of the D0 -- KK5 -- D4  brane intersection  depicted in Table~\ref{D0-KK-D4}.
 \begin{table}[ht]
	\begin{center}
		\begin{tabular}{| l | c | c | c | c| c | c| c | c| c | c |}
			\hline		    
			& 0 & 1 & 2 & 3 & 4 & 5 & 6 & 7 & 8 & 9 \\ \hline
			D0 & $\times$ &  & &  &  &  &  &   &   &   \\ \hline
				KK5 & $\times$ & $\times$ &$\times$  &$\times$  & $\times$ &$\times$   & $\text{ISO}$ &  &  &   \\ \hline
			D4 & $\times$ &  &  &  &  &  &  $\times$ & $\times$  &$\times$   &$\times$   \\ \hline
		\end{tabular} 
	\end{center}
	\caption{$\frac14$-BPS brane intersection underlying the Abelian T-dual solution. The D1-branes become D0-branes, the D3-branes become D4-branes wrapped on $y$ and the NS5-branes become KK5-monopoles with $y$ Taub-NUT direction.}   
	\label{D0-KK-D4}	
\end{table} 
Indeed, consider the following 10d background in type IIA supergravity,
\begin{equation}
   \begin{split}
    d s_{10}^2&=-H_{\mathrm{D}0}^{-1/2}\,H_{\mathrm{D}4}^{-1}\,d t^2+H_{\mathrm{D}0}^{1/2}\,H_{\mathrm{D}4}^{1/2}\,\left(d \rho^2+\rho^2\,d s^2_{S^4} \right)+H_{\mathrm{D}0}^{1/2}\,H_{\mathrm{D}4}^{-1/2}\,H_{\mathrm{KK}5}^{-1}\,\left(dy+  Q_{\mathrm{KK}5}\,\omega  \right)^2\,+\\
   &\phantom{=}\,+\,H_{\mathrm{D}0}^{1/2}\,H_{\mathrm{D}4}^{-1/2}\,H_{\mathrm{KK}5}\,\left(d r^2+r^2\,d s^2_{S^2} \right) \ ,\\
   C_{(1)}&=H_{\mathrm{D}0}^{-1}\, dt \ ,\qquad \, C_{(5)}=H_{\mathrm{KK}5}H_{\mathrm{D}4}^{-1}\, dt \wedge dy \wedge \mathrm{vol}_{\left(\mathbb{R}^3\right)}\ , \qquad e^{\Phi}=H_{\mathrm{D}0}^{3/4}\,H_{\mathrm{D}4}^{-1/4}\ ,
   \label{D0D4KK_metric}
   \end{split}
  \end{equation}
where, as in the type IIB case, we denoted by $\rho$ and $r$ the radial coordinates of $\mathbb{R}^5$ and $\mathbb{R}^3$ parameterised, respectively, by $(x^1,\dots,x^5)$ and $(x^7,x^8,x^9)$. It can be shown that \eqref{D0D4KK_metric} satisfies the equations of motion of type IIA supergravity if
$H_{\mathrm{D}0}=H_{\mathrm{D}0}(\rho,r)$, $H_{\mathrm{D}4}=H_{\mathrm{D}4}(\rho)$ and $H_{\mathrm{KK}5}=H_{\mathrm{KK}5}(r)$  satisfy
\begin{equation}
H_{\mathrm{D}0}=1+Q_{\mathrm{D}0}^{-2}\left(\frac{\pi Q_{\mathrm{D}4}}{\rho}\,+Q_{\mathrm{KK}5}\,r\right)\ ,\quad H_{\mathrm{D}4}=\frac{\pi Q_{\mathrm{D}4}}{\rho^3}
\quad \text{and}\quad H_{\mathrm{KK}5}=\frac{Q_{\mathrm{KK}5}}{r}\ .
\label{D0D4KK5_solution}
\end{equation}
As in the case of the D1 -- D3 -- NS5 brane setup we need to introduce new variables in order to make manifest the AdS geometry included into the near horizon. In particular, by introducing the $(\zeta,\, \alpha)$ coordinates as in \eqref{NH_coordinates} and taking the near horizon limit as $\zeta\,\rightarrow 0$, the background \eqref{IIAmetric} is recovered. We point out that D0 -- D4 brane intersections in the absence of NUT charge were originally considered in \cite{Cvetic:2000cj}. The regime of validity of the IIA solution, 
characterised by small string coupling and weakly curved limit, is when 
\begin{equation}
 N^{1/3}\ll k \ll N^3\,,\label{IIAregimeM5}
\end{equation}
where we have taken $k\equiv Q_{\mathrm{D}0}$ and $N\equiv Q_{\mathrm{D}4}$.
This is therefore the range of parameters where the superconformal quantum mechanics description is valid. 

\begin{figure}
\centering
\includegraphics[scale=0.8]{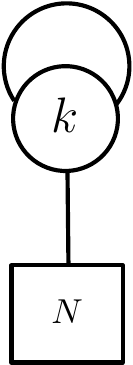}
\caption{Quiver describing the D0-D4 brane system.}
\label{longM5}
\end{figure}

The quantum mechanical system consisting on $k$ D0 and $N$ D4 branes \cite{Douglas:1996yp} was used in \cite{Berkooz:1996is,Aharony:1997th,Aharony:1997an} to describe the IR limit of the longitudinal M5-brane in M(atrix) theory. The D4-branes are M5-branes wrapped on the M-theory circle, and the D0-branes are M0-branes, that is, momentum charge along the eleventh direction. The D0 -- D4 system is thus equivalent to $N$ M5-branes compactified on a circle carrying $k$ units of momentum. The (2,0) six dimensional theory that describes the M5-branes is recovered from the quantum mechanics on the D0 -- D4 system in the large $N$ limit. 
The D4-branes break half of the supersymmetries of the D0-branes, and this is modelled  by adding some matter content to the $\mathrm{U}(k)$ quantum mechanics that describes them (see the discussion in \cite{Hanany:1997xc}). The way this is done is by adding $N$ fundamental hypermultiplets \cite{Douglas:1996yp}. The resulting quantum mechanics consists on a $\mathrm{U}(k)$ gauge theory with hypermultiplets in the adjoint representation and $N$ fundamentals. The corresponding quiver is depicted in Figure \ref{longM5}.
This theory preserves one quarter of the supersymmetries, and has a $\mathrm{SU}(2)\times \mathrm{SU}(2)\times \mathrm{SO}(5)$ global symmetry\footnote{The first $\mathrm{SU}(2)$ acts on the field in the adjoint representation, and the $\mathrm{SU}(2)\times \mathrm{SO}(5)$ part is the R-symmetry.}.
Note that our solution contains as well KK-monopoles, that arise from the NS5-branes present in type IIB. They are associated to the ALE singularity introduced by the orbifolding by $\mathbb{Z}_{k'}$. This orbifolding breaks the global symmetry to $\mathrm{SU}(2)\times \mathrm{SO}(5)$, but does not break any additional supersymmetries\footnote{This can be seen explicitly by observing that the projector acting on the Killing spinor of the brane setup \eqref{D0D4KK_metric} associated to the KK monopole can be obtained from those associated to the D0 and D4-branes.}.
We will propose in section \ref{field-theory} a quantum mechanics with eight supercharges that describes the effect of adding the KK-monopoles onto the D0 -- D4 system.

\subsection{Uplift to M-theory}\label{Mtheory}

In the last section we have been able to connect our type IIA picture with previously known constructions of SQM models arising from D0 -- D4 systems. It is worth noticing that the corresponding dual $\mathrm{AdS}_2$ backgrounds still suffer from the same non-compactness issue already discussed in type IIB. This behaviour is suggestive of the presence of deconstructed extra dimensions within the dual field theory. To support this intuition, we now show that such non-compact geometry is resolved by going to M-theory. In this picture, the non-compact direction will be hidden within a higher dimensional AdS geometry.

Indeed, uplifting the solution described by equations (\ref{IIAmetric})-(\ref{IIAF4}) we find
\begin{eqnarray}
&&d{s}_{11}^2=(\pi Q_{\mathrm{M}5})^{2/3}\left( 4\,ds^2_{\mathrm{AdS}_7/\mathbb{Z}_{k}\times \mathbb{Z}_{k'}}+ds^2_{S^4}\right)\ ,\label{AdS7_orbifoldI}\\
&&{G}_{(4)}=-3\,\pi Q_{\mathrm{M}5}\, {\rm vol}_{(S^4)} \ ,\label{AdS7_orbifoldII}
\end{eqnarray}
where, after the uplift, $k=Q_{\mathrm{M}0}$, $k'=Q_{\mathrm{KK}6}$, $Q_{\mathrm{D}4}=Q_{\mathrm{M}5}$, and
\begin{equation}
ds^2_{\mathrm{AdS}_7/\mathbb{Z}_k\times \mathbb{Z}_{k'}}=\frac14\,d\mu^2+\cosh^2{\frac{\mu}{2}}\,ds^2_{\mathrm{AdS}_3/\mathbb{Z}_{k}}+\sinh^2{\frac{\mu}{2}}\,ds^2_{S^3/\mathbb{Z}_{k'}}\ .
\end{equation}
Here we have redefined $\sin{\alpha}=\cosh^{-1}{\frac{\mu}{2}}$, with $\mu\in [0,\pi]$, and
\begin{equation}
\label{AdS3Z}
ds^2_{\mathrm{AdS}_3/\mathbb{Z}_{k}}=\frac14\left[\left(\frac{dz}{k}+\eta\right)^2+ds^2_{\mathrm{AdS}_2}\right]\qquad \text{with} \qquad d\eta={\rm vol}_{\left(\mathrm{AdS}_2\right)}\,.
\end{equation}
This background describes $N$ longitudinal M5-branes with $k$ momentum at an ALE singularity, associated to $k'$ KK-monopole charge. The momentum charge along the M-theory circle (D0 charge in type IIA) quotients $\mathrm{AdS}_3\rightarrow \mathrm{AdS}_3/\mathbb{Z}_{k}$, such that half of the supersymmetries of the $\mathrm{AdS}_7\times S^4$ background are broken. However, the amount of supersymmetry is not further reduced by the presence of the ALE singularity, which sends $S^3\rightarrow S^3/\mathbb{Z}_{k'}$. The ${\mathcal N}=8$ supersymmetries are preserved upon reduction to type IIA and, further, upon T-dualisation to type IIB. 

The brane intersection underlying the solution is depicted in Table \ref{M0-KK-M5}. It is described by the solution
\begin{table}[ht]
	\begin{center}
		\begin{tabular}{| l | c | c | c | c| c | c| c | c| c | c | c |}
			\hline		    
			& 0 & 1 & 2 & 3 & 4 & 5 & 6 & 7 & 8 & 9 & 10 \\ \hline
			M0 & $\times$ &  & &  &  &  &  &   &   &  & $\rightarrow$ \\ \hline
			KK6 & $\times$ & $\times$ &$\times$  &$\times$  & $\times$ & $\times$   & ISO &  &  & & $\times$   \\ \hline
			M5 & $\times$ &  &  &  &  &  &  $\times$ & $\times$  & $\times$  & $\times$ & $\times$ \\ \hline
		\end{tabular} 
	\end{center}
	\caption{$\frac14$-BPS brane intersection underlying the $\mathrm{AdS}_7/(\mathbb{Z}_{k}\times \mathbb{Z}_{k'})$ M-theory solution. The double orbifold reduces the supersymmetries of the M5-brane by a half.}   
	\label{M0-KK-M5}	
\end{table} 
\begin{equation}
   \begin{split}
    d s_{11}^2&=H_{\mathrm{M}5}^{-1/3}\,\left[-H_{\mathrm{M}0}^{-1}\,d t^2\,+\,H_{\mathrm{M}0}\,(dz\,+\,(H_{\mathrm{M}0}^{-1}-1)\,dt)^{2}\,+\,H_{\mathrm{KK}6}\,\left(d r^2+r^2\,d s^2_{S^2} \right) \right]\,+\\
   &\phantom{=}\,+\,H_{\mathrm{M}5}^{-1/3}\,H_{\mathrm{KK}6}^{-1}\,\left(dy+  Q_{\mathrm{KK}6}\,\omega  \right)^2+H_{\mathrm{M}5}^{2/3}\,\left(d \rho^2+\rho^2\,d s^2_{S^4} \right)\ ,\\
   A_{(6)} &= H_{\mathrm{KK}6}H_{\mathrm{M}5}^{-1}\, dt \wedge dy \wedge \mathrm{vol}_{\left(\mathbb{R}^3\right)}\,\wedge\,dz\ , 
   \label{M0M5KK_metric}
   \end{split}
  \end{equation}
where we have denoted by $z$ the coordinate parameterising the M-theory direction. The functions $H_{\mathrm{M}0}=H_{\mathrm{M}0}(\rho,r)$, $H_{\mathrm{M}5}=H_{\mathrm{M}5}(\rho)$ and $H_{\mathrm{KK}6}=H_{\mathrm{KK}6}(r)$ satisfying the equations of motion of 11d supergravity are given by
\begin{equation}
H_{\mathrm{M}0}=1+Q_{\mathrm{M}0}^{-2}\left(\frac{\pi Q_{\mathrm{M}5}}{\rho}\,+Q_{\mathrm{KK}6}\,r\right)\ ,\quad H_{\mathrm{M}5}=\frac{\pi Q_{\mathrm{M}5}}{\rho^3}
\quad \text{and}\quad H_{\mathrm{KK}6}=\frac{Q_{\mathrm{KK}6}}{r}\ .
\label{M0M5KK_solution}
\end{equation}
The integration constants are now interpreted as quantised momentum, M5 and NUT charges, respectively.
The introduction of the $(\zeta,\, \alpha)$ coordinates as in \eqref{NH_coordinates} gives rise this time to a locally AdS$_7$ geometry in the $\zeta\,\rightarrow 0$ limit. We point out that M0 -- M5 brane intersections in the absence of NUT charge were originally considered in \cite{Cvetic:2000cj}.

\vspace{2mm}
\noindent \textbf{Holographic free energy:} The free energy of the 6d CFT dual to the solution can be computed holographically from the worldvolume of the M5-branes. In this calculation the singularity found in the type II descriptions is  absorbed in the (infinite) worldvolume of the M5-branes. For $N$ M5-branes wrapped on $\mathrm{AdS}_3/\mathbb{Z}_{k}\times S^3/\mathbb{Z}_{k'}$ the effective number of degrees of freedom evaluates to
\begin{equation}
S_{\mathrm{M}5}=-T_5\int d^6\xi \sqrt{{\rm det}\tilde{g}}=4\pi^5\sinh^3{\mu}\,{\rm Vol}(\mathrm{AdS}_2) \ \frac{N^3}{kk'} \ ,\label{M5dof}
\end{equation}
where ${\tilde g}$ is the induced metric and we have used that $T_5=Q_{\mathrm{M}5}=N$. This result reproduces the scaling in \eqref{FHol_IIB}, though within the context of a 6d effective description. This corroborates our  proposal that our 1d CFT is UV completed by a 6d CFT,  dual to the background \eqref{AdS7_orbifoldI}-\eqref{AdS7_orbifoldII}. In order to further clarify our proposal we now discuss 
in more detail the regime of validity of the type II picture against the M-theory one, and their interplay. 

From the point of view of the type II backgrounds, we pointed out that the divergence of the free energy is due to an infinite amount of integrated $F_{(7)}$ ($F_{(8)}$) flux defining magnetically the D1 (D0) brane charge. On the other hand, the aforementioned charge computed electrically is given by $k$, i.e.
\begin{equation}
 Q_{\mathrm{D}1}^{\text{el}}=Q_{\mathrm{D}0}^{\text{el}}\sim k \qquad \text{and} \qquad Q_{\mathrm{D}1}^{\text{mag}}=Q_{\mathrm{D}0}^{\text{mag}}\sim \frac{N^3}{k^2 k'}\, \cot^4 \epsilon\,,
\end{equation}
where we have neglected order 1 factors. The 1d theory is consistently described by the physics of D1 or D0-branes only when it is possible to impose $Q_{\mathrm{D}1}^{\text{el}}=Q_{\mathrm{D}1}^{\text{mag}}$, $Q_{\mathrm{D}0}^{\text{el}}=Q_{\mathrm{D}0}^{\text{mag}}$. This condition forces one to introduce a charge dependent cut-off
\begin{equation}
 \cot^4\epsilon\sim \frac{k^3 k'}{N^3}\,,
\end{equation}
which stays consistently finite within the IIA regime obtained in \eqref{IIAregimeM5}. In this limit the free energy given by (\ref{FHol_IIB}) has the scaling behaviour
\begin{equation}
 {\cal F}_{\mathrm{hol}}\sim k^2\ ,
\label{FHol_IIA_2}
\end{equation}
which can be reproduced by evaluating the DBI action of $k$ D0 or D1 brane probes in the corresponding $\mathrm{AdS}_2\times S^4$ backgrounds. Remarkably, this is the scaling of the free energy obtained in \cite{Hartman:2008dq} for generic $\mathrm{AdS}_2$ solutions with non-vanishing electric field.

Conversely, in a limit where the cut-off diverges while remaining charge independent, one can no longer enforce that the charges computed electrically and magnetically coincide. As a consequence an M-theory regime emerges, and a way to see this is that the $\frac{N^3}{k k'}$ factor in the type II free energy will not compensate the unbounded growth of the $\cot^4\epsilon$ cut-off, at least not within a good trustworthy M-theory description\footnote{Note that we better stick to the $k\ll N^3$ regime if we want to stay within a weakly curved approximation that allows us to neglect higher derivatives suppressed by powers of the 11d Planck length $\ell_{11}$.}.
Consistently, in this situation the free energy given by (\ref{FHol_IIB}) reproduces the typical scaling of the free energy of the 6d theory \eqref{M5dof} obtained by counting M5-brane worldvolume degrees of freedom.

\section{Field Theory Description} \label{field-theory}

\begin{figure}
\centering
\includegraphics[scale=0.6]{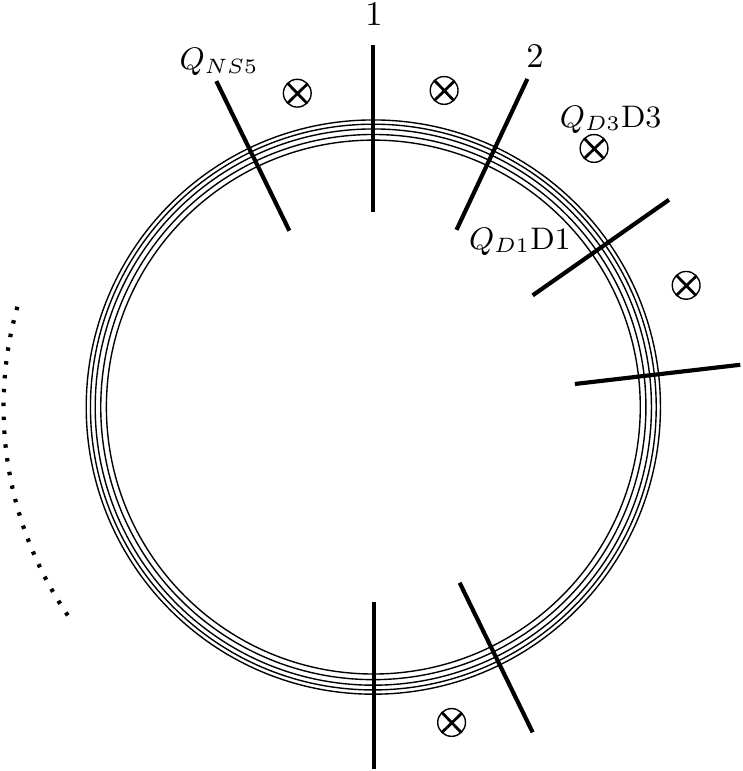}
\caption{Circular quiver describing longitudinal M5-branes with momentum charge probing an ALE singularity.}
\label{circular-quiver}
\end{figure}

In this section we propose a quantum mechanics with 8 supercharges that describes the M0-KK-M5 brane system. We consider its realisation in type IIB, depicted in Table \ref{D1-NS5-D3}. This brane set-up can be related by T-dualities to the D3-NS5-D5 brane system first discussed in \cite{Hanany:1996ie}. The resulting quantum mechanics living in the D1-branes contains 8 supercharges, and has an $\mathrm{SU}(2)\times \mathrm{SO}(5)$ global symmetry, which is an R-symmetry. The matter content consists on a $\mathrm{U}(Q_{\mathrm{D}1})^{Q_{\mathrm{NS}5}}$ gauge group with $Q_{\mathrm{NS}5}$ hypermultiplets transforming in the bifundamental of the two adjacent $\mathrm{U}(Q_{\mathrm{D}1})$ gauge groups. On top of this there are $Q_{\mathrm{NS}5}$ hypermultiplets in the fundamental of $\mathrm{U}(Q_{\mathrm{D}3})$. The corresponding circular quiver is depicted in Figure \ref{circular-quiver}. 

In the absence of D3-branes (M5-branes in eleven dimensions) the quiver defines the quantum mechanics with 8 supercharges that describes the KK-monopole in eleven dimensions  \cite{Hanany:1997xc}. In the opposite situation, when there is just one NS5-brane, and therefore no KK-monopoles\footnote{Recall that one KK-monopole is indistinguishable from no KK-monopoles.}, the bifundamental hypermultiplet becomes an adjoint hypermultiplet, and the 1d theory realises the quantum mechanics that describes the longitudinal M5-brane \cite{Aharony:1997th,Aharony:1997an}. The corresponding quiver is depicted in Figure \ref{longM5}. This QM has $\mathrm{SO}(4)\times \mathrm{SO}(5)$ global symmetries, with $\mathrm{SU}(2)\times \mathrm{SO}(5)$ being R-symmetries and the additional $\mathrm{SU}(2)$ being a global symmetry acting on the adjoint hypermultiplet. 

 Even if this global symmetry is not manifest in the type IIB description, it becomes manifest  in the type IIA set-up, where the configuration becomes the one depicted in Table \ref{D0-KK-D4}, for zero  KK-monopoles. Our proposal is thus consistent with the descriptions in the literature of KK-monopoles  \cite{Hanany:1997xc} and longitudinal 5-branes \cite{Aharony:1997th,Aharony:1997an} in eleven dimensions. Note that even if these descriptions were formulated in the context of M(atrix) theory, they can be extended to the case in which the eleventh direction is not a light-cone but an ordinary spatial direction  \cite{Sen:1997we,Seiberg:1997ad}. The situation in which both KK-monopoles and M5-branes are absent can also be considered. In this case 
the quantum mechanics consists on a $\mathrm{U}(Q_{\mathrm{D}1})$ gauge group coupled to a hypermultiplet in the adjoint representation. When the mass of the adjoint is zero the number of supersymmetries is enhanced from 8 to 16, and the QM becomes just the BFSS proposal \cite{Banks:1996vh}. 

The previous construction should be extendable to non-compact D1-NS5-D3 brane systems defining linear quiver quantum mechanics with 8 supercharges.
These quantum mechanics would flow to superconformal quantum mechanics in the IR, that would be holographically dual to solutions in the class in \cite{DHoker:2007mci} for which the 2d Riemann surface is non-compact. The one to one mapping found in \cite{Assel:2011xz} between 3d SCFTs described by linear quivers and $\mathrm{AdS}_4\times S^2\times S^2\times \Sigma_2$ solutions in the classification in \cite{DHoker:2007zhm,DHoker:2007hhe}, for $\Sigma_2$ an infinite strip\footnote{Or upper half-plane.}, suggests that a similar mapping should be possible between 1d SCFTs and $\mathrm{AdS}_2\times S^4\times S^2\times \Sigma_2$ solutions for $\Sigma_2$ an infinite strip, upon analytical continuation. We will discuss in section \ref{NATD} a particular $\mathrm{AdS}_2\times S^4\times S^2$ solution to type IIB in the class in \cite{DHoker:2007mci} associated to a linear quiver quantum mechanics.

\section{$\mathrm{AdS}_2$ Solutions from D0 -- F1 Branes}\label{D0-D1}

In the previous sections we have worked out the relation between a certain non-compact ten dimensional background involving $\mathrm{AdS}_2$ and a 1d superconformal quantum mechanics  capturing the physics of an IR phase of the worldvolume theory of an M5-brane stack. From the supergravity viewpoint an important piece of evidence for this was provided by the emergence of a locally $\mathrm{AdS}_7$ geometry when constructing the corresponding M-theory lift.

The aim of this section is to provide a similar mechanism, this time within the context of the worldvolume theory of a stack of M2-branes instead.
To this end, we show that the $\mathrm{AdS}_2\times S^7\times I_\alpha$ solution of massless type IIA supergravity constructed in \cite{Dibitetto:2018gbk} (suffering from a similar non-compactness problem in the $\alpha$-direction) is dual to the superconformal quantum mechanics with 8 supercharges that describes M2-branes with momentum charge. We start with the solution constructed in \cite{Dibitetto:2018gbk}:
\begin{eqnarray}
&&ds_{10}^2=\frac{\pi\,\ell^2}{8\sqrt{2}\, \sin^3{\alpha}}\left( ds^2_{\mathrm{AdS}_2}+4d\alpha^2+16\,\sin^2{\alpha}\,ds^2_{S^7}\right)\ , \label{AdS2S7_metric}\\
&&e^\Phi=\frac{\pi^{1/2}}{2^{5/4}}\,\frac{Q_{\mathrm{F}1}^{1/4}}{Q_{\mathrm{D}0}^{3/2}}\,\sin^{-3/2}{\alpha} \ ,\\
&& F_{(2)}=Q_{\mathrm{D}0} \,{\rm vol}_{\left(\mathrm{AdS}_2\right)}\ ,\\
&&H_{(3)}=-\frac{3\,\pi}{8\sqrt{2}}\,\frac{Q_{\mathrm{F}1}^{1/2}}{Q_{\mathrm{D}0}}\,\sin^{-4}{\alpha}\,{\rm vol}_{\left(\mathrm{AdS}_2\right)}\wedge d\alpha\ , \label{AdS2S7_flux}
\end{eqnarray}
where $\ell^2=Q_{\mathrm{F}1}^{1/2}Q_{\mathrm{D}0}^{-1}$, while $Q_{\mathrm{D}0}=k$ and $Q_{\mathrm{F}1}=N$ are quantised in string units.
This solution is obtained as the near horizon limit of the (semilocalised) D0 -- F1 system  shown in Table \ref{D0-F1} \cite{Cvetic:2000cj}. 
\begin{table}[ht]
	\begin{center}
		\begin{tabular}{| l | c | c | c | c| c | c| c | c| c | c |}
			\hline		    
			& 0 & 1 & 2 & 3 & 4 & 5 & 6 & 7 & 8 & 9  \\ \hline
			D0 & $\times$ &  & &  &  &  &  &   &   &   \\ \hline
			F1 & $\times$ & $\times$ &  &  &  &   &  &  &  &    \\ \hline
		\end{tabular} 
	\end{center}
	\caption{$\frac14$-BPS brane intersection underlying the $\mathrm{AdS}_2\times S^7\times I_{\alpha}$ solution. This brane setup is a particular example in the class of \cite{Imamura:2001cr}.}   
	\label{D0-F1}	
\end{table} 
Specifically, the AdS radial coordinate turns out to be a non-trivial combination of $x^1$ and the radial coordinate of the transverse $\mathbb{R}^8$ parameterised by $\left(x^3,\,\dots,\,x^9\right)$ in Table \ref{D0-F1}.
The $\mathrm{SO}(2,1)\times \mathrm{SO}(8)$ isometries of the solution match the bosonic subgroup of the $\mathrm{OSp}(8|2)$ supergroup (see Table \ref{groups}). Thus, this should be the supergroup associated to the superconformal quantum mechanics dual to the solution. Note that the $\mathrm{OSp}(8|2)$ supergroup allows as well for a bosonic subgroup $\mathrm{SO}(2, 6)\times \mathrm{SO}(3)$. This is the group of isometries of the $\mathrm{AdS}_7\times S^2$ solutions to massive Type IIA supergravity constructed in \cite{Apruzzi:2013yva}, which, as shown in \cite{Dibitetto:2018gbk}, are related to the $\mathrm{AdS}_2\times S^7$ solutions upon analytical continuation.

\vspace{2mm}
\noindent \textbf{Holographic Free Energy:}
Given the non-compactness of the supergravity background shown in \eqref{AdS2S7_metric}-\eqref{AdS2S7_flux}, we do expect once again to find a divergent result when holographically evaluating the effective number of degrees of freedom of the dual field theory. 
We point out that such a computation holds in the regime characterised by small string coupling and weakly curved limit for the background \eqref{AdS2S7_metric}. This amounts to requiring
\begin{equation}
 N^{1/6}\ll k \ll N^6\,.\label{IIAregimeM2}
\end{equation}
By specifying the formulas \eqref{Hhat} and \eqref{centralx} to this case, we obtain
\begin{equation}
\sqrt{{\hat H}}=\frac{\pi^{13}}{3\cdot 2^{15/2}}\frac{N^{3/2}}{k}\cot{\epsilon}\,,
\end{equation}
where we have regularised the divergence due to the non-compactness of $I_\alpha$ with a hard cut-off $\epsilon$. Substituting this into equation (\ref{centralx}) we find
\begin{equation}
{\cal F}_{\rm hol}=\frac{\pi^7}{2^{21/2}}\frac{N^{3/2}}{k}\cot{\epsilon} \ .\label{F_hol_D0F1}
\end{equation}
The divergence of the free energy suggests that the ultimate dual desciption in the UV should be given by a higher dimensional CFT rather than a superconformal quantum mechanics. Nevertheless, in analogy to what happened for the previous case of the M5-branes, we do expect to retain a 1d effective description valid in the IR. In the following we will argue, by looking at the uplift to eleven dimensions, that the higher dimensional UV completion should be the 3d CFT associated to M2-branes, supplemented with extra momentum states.

\subsection{Uplift to M-theory}

The D0 -- F1 brane intersection illustrated above has a natural interpretation in M-theory in terms of longitudinal M2-branes, i.e. M2-branes carrying momentum charge. The corresponding supergravity background is semilocalised \cite{Cvetic:2000cj}, its near horizon geometry being
\begin{eqnarray}
&&d{ s}_{11}^2=L^2 \left( ds^2_{\mathrm{AdS}_4/\mathbb{Z}_k}+4 ds^2_{S^7}\right) \\
&&{ G}_{(4)}=3L^3 {\rm vol}_{\mathrm{AdS}_4/\mathbb{Z}_k}
\end{eqnarray}
where $L^6$ is proportional to the quantised M2-brane charge $Q_{\mathrm{M}2}=Q_{\mathrm{F}1}=N$ through the relation $L^6=2^{-1}\pi^2Q_{\mathrm{F}1}$. The metric of $\mathrm{AdS}_4/\mathbb{Z}_k$ is parametrised as
\begin{equation}
ds^2_{\mathrm{AdS}_4/\mathbb{Z}_k}=d\mu^2+\cosh^2{\mu}\,ds^2_{\mathrm{AdS}_3/\mathbb{Z}_k}
\end{equation}
and the $\mu$ coordinate relates through $\sin{\alpha}=\cosh^{-1}{\mu}$ to the IIA warping coordinate $\alpha$. The metric $ds^2_{\mathrm{AdS}_3/\mathbb{Z}_k}$ reads as in (\ref{AdS3Z}).
The above solution describes $N$ M2-branes with $k$ momentum charge. This sends $\mathrm{AdS}_3\rightarrow \mathrm{AdS}_3/\mathbb{Z}_k$, such that half of the supersymmetries of the $\mathrm{AdS}_4\times S^7$ background are broken.

The free energy can be computed from the worldvolume of the M2-branes. In this calculation the singularity found in the type IIA description is absorbed in the (infinite) worldvolume of the M2-branes. For $N$ M2-branes wrapped on $\mathrm{AdS}_3/\mathbb{Z}_k$ we find
\begin{equation}
S_{\mathrm{M}2}=-T_2\int d^3\xi \sqrt{{\rm det}\tilde g}= \frac{\pi^2}{2^{5/2}} \frac{N^{3/2}}{k}\cosh^3\mu{\rm Vol}\left( \mathrm{AdS}_2 \right)\,, \label{M2free}
\end{equation}
where we have used that $T_2=Q_{\mathrm{M}2}=N$. This result is in agreement with the type IIA calculation \eqref{F_hol_D0F1}. We may now proceed with a similar analysis to the one done in section \ref{Mtheory}.

The divergence of the free energy \eqref{F_hol_D0F1} is again due to an infinite amount of integrated $F_{(8)}$ flux, from which the D0-brane charge is computed magnetically. In particular we have that
\begin{equation}
 Q_{\mathrm{D}0}^{\text{el}}\sim k \qquad \text{and} \qquad Q_{\mathrm{D}0}^{\text{mag}}\sim \frac{N^{3/2}}{k^2}\, \cot \epsilon\,,
\end{equation}
where we have neglected order 1 factors. By imposing that $Q_{\mathrm{D}0}^{\text{el}}=Q_{\mathrm{D}0}^{\text{mag}}$, one again finds a charge dependence for the cut-off
\begin{equation}
 \cot\epsilon\sim \frac{k^3}{N^{3/2}}\,,\label{cutoffM2}
\end{equation}
which stays consistently finite within the IIA regime obtained in \eqref{IIAregimeM2}. This situation turns out to describe again a D0-brane quantum mechanics. By virtue of \eqref{cutoffM2} the free energy in \eqref{F_hol_D0F1} evaluates to
\begin{equation}
 {\cal F}_{\mathrm{hol}}\sim k^2\ .
\end{equation}
Remarkably the last result manifests the same universal scaling behaviour for $\mathrm{AdS}_2$ solutions found in  \cite{Hartman:2008dq},  and reproduced  when studying the IR phase of longitudinal M5-branes in section 2.3. Moreover, also in this case the same outcome is reproduced by directly evaluating the DBI action of $k$ D0-brane probes in $\mathrm{AdS}_2\times S^7\times I_\alpha$. This analogy confirms that in both situations the IR descriptions of worldvolume theories of M-branes may be given in terms of a D0-brane quantum mechanics. Reversely, in the regime in which the M-theory description is valid the charge dependence of the cut-off in \eqref{cutoffM2} does not make sense, and one has to resort to the eleven dimensional description, given by \eqref{M2free}. Consistently, in this situation the free energy reproduces the typical scaling of the free energy of the 3d theory associated to M2-branes.

Our previous result is in agreement with the M(atrix) theory description of longitudinal M2-branes \cite{Aharony:1996bh}. In this framework computations such as the potential between a pair of branes and between 
branes and gravitons were shown to agree with the supergravity results. Note that, as in section 2.3, the momentum direction is not a light-cone direction in this case but an ordinary spatial direction, with both of them being related through the Sen-Seiberg limit  \cite{Sen:1997we,Seiberg:1997ad}.

\section{$\mathrm{AdS}_2\times S^4\times S^2\times \Sigma_2$ Solutions and Linear Quivers}\label{NATD}

As we discussed at the end of section \ref{field-theory}, it should be possible to construct more general type IIB $\mathrm{AdS}_2\times S^4\times S^2$ solutions dual to superconformal quantum mechanics described by linear quivers. In this section we provide one such example through non-Abelian T-duality (NATD)   \cite{delaOssa:1992vci,Sfetsos:2010uq}. We argue that the resulting solution is associated to a superconformal quantum mechanics described by a linear quiver of gauge groups with increasing ranks. 

Our starting point is the $\mathrm{AdS}_2\times S^4\times S^3/\mathbb{Z}_{k'}$ solution to type IIA discussed in section \ref{ATD}. We consider the simpler case in which $k'=1$ and
perform a non-Abelian T-duality transformation 
with respect to a freely acting $\mathrm{SU}(2)$ on the  $S^3$. The resulting solution fits locally in the class of solutions constructed in \cite{DHoker:2007mci} for $\Sigma_2$ an infinite strip and, as the solutions in this class, it is related through analytical continuation to the $\mathrm{AdS}_4\times S^2\times S^2\times \Sigma_2$ solutions in \cite{DHoker:2007zhm,DHoker:2007hhe}. The specific 
$\mathrm{AdS}_4\times S^2\times S^2\times \Sigma_2$ solution involved  is dual to a 3d CFT in the class of \cite{Gaiotto:2008ak}, described by a linear quiver of gauge groups with increasing ranks. This strongly suggests that a similar description should be at play for the $\mathrm{AdS}_2$ solution.
As in similar examples in which non-Abelian T-duality has been applied to holography \cite{Lozano:2016kum,Lozano:2016wrs,Lozano:2017ole,Itsios:2017cew,Lozano:2018pcp,Lozano:2019zvg,Lozano:2019ywa}, the solution constructed through non-Abelian T-duality would not be describing the same physics as the type IIA seed solution, which would be consistent with the fact that non-Abelian T-duality has not been proved to be a string theory symmetry  \cite{Giveon:1993ai,Alvarez:1993qi}. 

We start by describing the non-Abelian T-dual solution. It is given by
\begin{eqnarray}
&&ds^2_{10}=\frac{\ell^2}{\sin^3{\alpha}}\left(ds^2_{\mathrm{AdS}_2}+4d\alpha^2+\sin^2{\alpha}ds^2_{S^4}\right)+\frac{\ell^2 \cos^2{\alpha}\sin^3{\alpha}}{\Delta}r^2 ds^2_{S_2}+\frac{\sin^3{\alpha}}{\ell^2 \cos^2{\alpha}} dr^2 \,,  \label{IIBNATD} \\
&&e^\Phi=\frac{\sin^3{\alpha}}{Q_{\mathrm{D}0}\cos{\alpha}\sqrt{\Delta}}\,, \\
&&B_{(2)}=-\frac{\sin^6{\alpha}}{\Delta}r^3{\rm vol}_{\left(S^2\right)}\,, \\
&&F_{(3)}=-Q_{\mathrm{D}0}\left(r dr+6\ell^4\,\frac{\cos^3{\alpha}}{\sin^7{\alpha}}d\alpha\right)\wedge{\rm vol}_{\left(\mathrm{AdS}_2\right)}\,, \\
&&F_{(5)}=-\pi Q_{\mathrm{D}4}\left(3r dr+2\ell^4\,\frac{\cos^3{\alpha}}{\sin^5{\alpha}}d\alpha \right)\wedge{\rm vol}_{\left(S_4\right)}\nonumber\\&&\;\;\;\;\;\;\;\;\;-\frac{\ell^2\pi Q_{\mathrm{D}4}\cos^3{\alpha}}{\Delta}r^2\left(\cos{\alpha} dr-6\frac{r}{\sin{\alpha}}d\alpha\right)\wedge{\rm vol}_{\left(\mathrm{AdS}_2\right)}\wedge{\rm vol}_{\left(S_2\right)}\,, 
\end{eqnarray}
where we have introduced
\begin{eqnarray}
\Delta=\ell^4\cos^4{\alpha}+r^2\sin^6{\alpha}
\end{eqnarray}
and $\ell^2=\pi Q_{\mathrm{D}4}/Q_{\mathrm{D}0}$, in terms of the quantised charges of the type IIA solution prior to the dualisation.
It can be seen that it fits in the general classification in \cite{DHoker:2007mci}, for the choice of harmonic functions 
\begin{eqnarray}
h_1=\frac{\ell^2}{2}Q_{\mathrm{D}0}\gamma\;r	,\;\;\;\;\;\;\;\;\;\;\;\;\;\;\;\;\;\;\;\;\;\;\;\;\;\;\;\;\;\;\;\;\;\;\;\;\;\;\;\;\;&&\;\;\;\;\;h_2=\frac{\ell^2}{2}(1+\gamma),\nonumber\\
h_1^D=-\frac{\ell^2Q_{\mathrm{D}0}(1-2\gamma^2)+2(2c_0+Q_{\mathrm{D}0}r^2)}{8},\;\;\;\;\;&&\;\;\;\;\;h_2^{D}=\frac{r-n\pi}{2},
\end{eqnarray}
where we introduced $\gamma=\tan^{-2}\alpha$, $d(c_0)=0$ and $n$ is associated to large gauge transformations of the $B_{(2)}$-field (see below).

\vspace{2mm}
\noindent \textbf{Quantised charges:} The quantised charges are obtained from the Page fluxes
\begin{eqnarray}
	&&\hat{F}_{(3)}=F_{(3)}\,,\\
	&&\hat{F}_{(5)}=-\pi Q_{\mathrm{D}4}\left(3rdr+2\ell^4\frac{\cos^3{\alpha}}{\sin^5{\alpha}} d\alpha \right)\wedge {\rm vol}_{\left(S_4\right)}-Q_{\mathrm{D}0}r^2dr\wedge{\rm vol}_{\left(\mathrm{AdS}_2\right)}\wedge{\rm vol}_{\left(S^2\right)}\,,\\
	&&\hat{F}_{(7)}=-3\pi Q_{\mathrm{D}4}r^2 dr\wedge{\rm vol}_{\left(S_4\right)}\wedge{\rm vol}_{\left(S^2\right)}.
\end{eqnarray}
From them we can infer that the D0-brane of the type IIA solution gets mapped onto a D1-brane extended in $r$ plus a D3-brane wrapped in $r\times S^2$. As is common through non-Abelian T-duality, the D3-brane carries dielectric charge due to D1-branes opening up onto an $S^2$, because of non-vanishing $B_{(2)}$-charge. We will use D3-branes as colour branes for our solution. In turn, the D4-branes wrapped on $\mathrm{AdS}_2\times S^3$ are mapped onto D1-branes wrapped on $\mathrm{AdS}_2$ and D3-branes wrapped on $\mathrm{AdS}_2\times S^2$. These are non-compact branes that play the role of flavour branes. As we show below, they do not carry independent charges either. We will choose the D1-branes to play the role of flavour branes of our configuration. With these choices the brane set-up is read from a D1 -- NS5 -- D3 brane configuration, as the (Abelian T-dual) solution discussed in section 2.1. One can then easily check that the solution in section 2.1 arises in the $r\rightarrow\infty$ limit, as it should be the case \cite{Macpherson:2015tka,Lozano:2016kum}. 

For the correct computation of the quantised charges it is necessary to include large gauge transformations of the $B_{(2)}$-field. The 3d internal space spanned by $(r,S^2)$ is topologically $I\times S^2$, so there are large gauge transformation associated to the $S^2$. Given the form of the $B_{(2)}$-field we need to divide the $r$ direction into $[r_n,r_{n+1})$ intervals, for $r_n$ satisfying that
$B_{(2)}(r_n)=n\pi {\rm vol}_{S^2}$.
Given this, one unit of NS5-brane charge is being created at each interval, and a large gauge transformation of parameter $n$ needs to be performed to enforce that $\frac{1}{4\pi^2}\oint_{S^2}B_{(2)}\in [0,1)$. Due to the non-compactness of the internal space, the D3 and D1 colour charges computed from the ${\hat F}_{(5)}$ and ${\hat F}_{(7)}$ magnetic Page fluxes are infinite. As in previous sections we have to compute the charges as electric charges, from the  ${\hat F}_{(5)}$ and ${\hat F}_{(3)}$ electric Page fluxes, respectively. We then find
\begin{equation}
Q_{\mathrm{D}1}=\frac{\pi^2}{2}(2n+1)\, Q_{\mathrm{D}0}\, \qquad Q_{\mathrm{D}3}=2\pi^4 (n+\frac23)\, Q_{\mathrm{D}0}\, ,
\end{equation}
in each of the $[r_n,r_{n+1})$ intervals. In turn,
the flavour branes can be computed from the magnetic ${\hat F}_{(7)}$ and ${\hat F}_{(5)}$ fluxes, leading to
\begin{equation}
{\tilde Q}_{\mathrm{D}1}=\frac{\pi}{4}(n+\frac23)\, Q_{\mathrm{D}4}\, , \qquad {\tilde Q}_{\mathrm{D}3}=\frac{\pi}{4}(2n+1)\, Q_{\mathrm{D}4}\, .
\end{equation}
We will choose $Q_{\mathrm{D}3}$, ${\tilde Q}_{\mathrm{D}1}$ and $Q_{\mathrm{NS}5}$ as independent charges of our configuration in a given $[r_n,r_{n+1})$ interval.

\vspace{2mm}
\noindent \textbf{Brane singularities:} The solution \eqref{IIBNATD} has two singularities at the boundaries of the range of $\alpha$. The first lies at $\alpha=0$. After introducing $\beta=\alpha^{-2}$ one finds
\begin{eqnarray}
&&ds_{10}^2\sim \ell^2\beta^{3/2}ds^2_{\mathrm{AdS}_2}+\ell^2\beta^{-3/2}\left(d\beta^2+\beta^2ds^2_{S^4}+\ell^{-4}\left(dr^2+r^2ds^2_{S_2}\right)\right)\,,\\
&&e^\Phi\sim\ell^{-2}Q_{\mathrm{D}0}^{-1}\;\beta^{-3/2}.
\end{eqnarray}
This is the behaviour of D1-branes with worldvolume AdS$_2$, localised at the origin of $\mathbb{R}^5$ and smeared on $S^2\times I$.
The second singularity at $\alpha=\frac{\pi}{2}$ is almost identical to the behaviour at $\alpha=\pi/2$ found in section \ref{TypeIIB_Picture}. We have
\begin{eqnarray}
&&ds_{10}^2\sim \ell^2\left(ds^2_{\mathrm{AdS}_2}+ds^2_{S^4}\right)+\ell^2\beta^{-1}\left(d\beta^2+\beta^2ds^2_{S_2}+\ell^{-4}dr^2\right)\,,\\
&&e^\Phi\sim r^{-1}Q_{\mathrm{D}0}^{-1}\;\beta^{-1/2}\,,
\end{eqnarray}
where $\beta=(\frac{\pi}{2}-\alpha)^2$. This reproduces the behaviour of NS5-branes with worldvolume on AdS$_2\times S^4$, localised at the origin of $\mathbb{R}^3$ and smeared on the $r$ coordinate.

\begin{figure}
\centering
\includegraphics[scale=0.55]{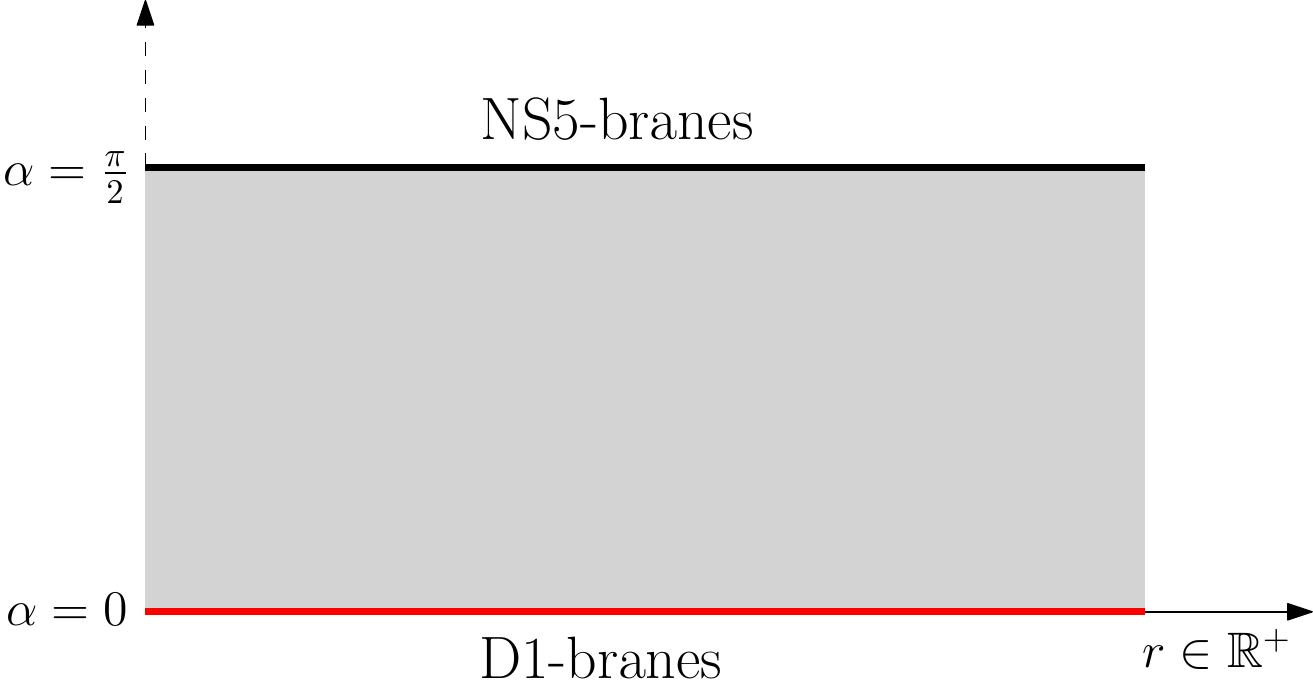}
\caption{Infinite strip associated to the NATD solution.}
\label{Riemann2}
\end{figure}

The Riemann surface associated to the solution is the infinite strip depicted in Figure \ref{Riemann2}. D1-branes are smeared over the lower boundary at $\alpha=0$ and NS5-branes are smeared on the upper boundary at $\alpha=\frac{\pi}{2}$. The strip topology follows from the unboundedness of the $r$ direction.

\begin{figure}
\centering
\includegraphics[scale=0.6]{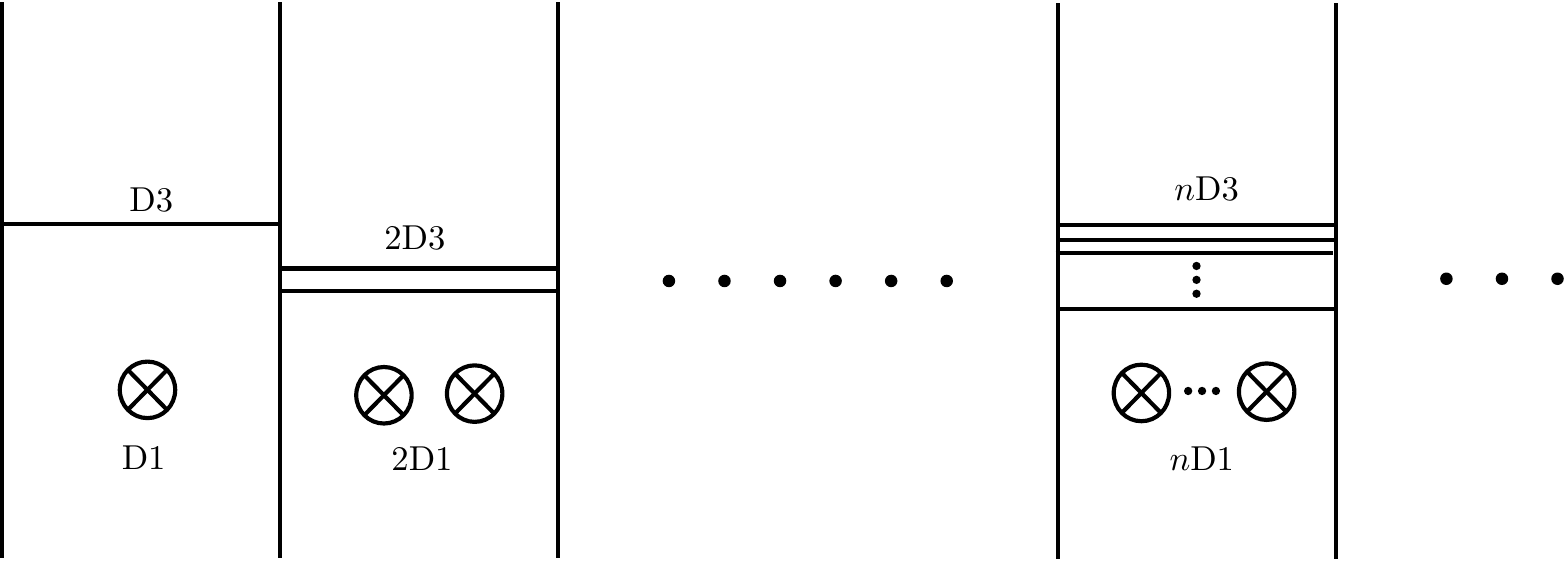}
\caption{Brane set-up underlying the non-Abelian T-dual solution. The D3-branes are wrapped on $S^2$, such that the effective field theory living in them is one dimensional at low energies.}
\label{brane-set-up}
\end{figure}

\vspace{2mm}
\noindent \textbf{Brane set-up:} The previous analysis is consistent with the brane set-up depicted in Figure \ref{brane-set-up}.
The quantum mechanics would live on D3-branes, carrying $Q_{\mathrm{D}3}$ charge, wrapped on $S^2$ and stretched between NS5-branes wrapped on $\mathrm{AdS}_2\times S^4$ and located at $\alpha=\pi/2$, $r=r_n$. D1-branes, carrying ${\tilde Q}_{\mathrm{D}1}$ charge, wrapped on $\mathrm{AdS}_2$ and located at $\alpha=0$ would provide additional flavour charges. The numbers of colour D3 and flavour D1 branes increase, together with the number of NS5-branes, as we move in $r$, which plays in this sense the role of field theory direction. The number of D3-branes increases in one unit in units of $2\pi^4 Q_{\mathrm{D}0}$, while the number of D1-branes increases in units of $\pi^2 Q_{\mathrm{D}0}$. 

The brane set-up depicted in Figure \ref{brane-set-up} suggests that it should be possible to construct superconformal quantum mechanics with 8 supercharges built out of linear quivers constructed from gauge groups of increasing ranks and increasing number of flavours. One would of course have to give a meaning to the $r$ non-compact direction, inherent to solutions constructed through non-Abelian T-duality, and to the non-compactness along the $\alpha$-direction, inherited from the type IIA solution. We will find an interesting relation between the corresponding divergences below when we address the free energy calculation. 
A more careful study of linear quivers superconformal quantum mechanics will however be left for future analysis.

\vspace{2mm}
\noindent \textbf{Free energy:} As in section 2 we can compute the free energy using that
\begin{equation}
{\cal F}=\frac{3}{G_N}\sqrt{{\hat H}}
\end{equation}
In $\sqrt{{\hat H}}$ the volume of the internal space appears, which is divergent due to the non-compactness of both the $r$ and $\alpha$ directions. We regularise it introducing an IR cut-off in $r$ ($P$), as well as a UV cut-off in $\alpha$ ($\epsilon$), and we get
\begin{eqnarray}
&&\sqrt{{\hat H}}=\frac{2^6\pi^6}{3}\frac{Q_{\mathrm{D}4}^3}{Q_{\mathrm{D}0}}\int_{0}^{P}r^2 dr\int_\epsilon^{\pi/2}\frac{\cos^3{\alpha}}{\sin^5{\alpha}}d\alpha,\\
&&{\cal F}=\frac{2}{3}\frac{Q_{\mathrm{D}4}^{3}}{Q_{\mathrm{D}0}} P^3\cot^4{\epsilon}
\end{eqnarray}
In terms of the quantised charges of the non-Abelian T-dual solution this can be expressed as
\begin{equation}
\label{FHol_NATD}
{\cal F}\sim\frac{{\tilde Q}_{\mathrm{D}1}^3}{Q_{\mathrm{D}3}Q_{\mathrm{NS}5}}\cot^4{\epsilon}\, ,
\end{equation}
where ${\tilde Q}_{\mathrm{D}1}$ here states for the total flavour D1-brane charge in the $r\in [0,P]$ interval, $Q_{\mathrm{D}3}$ for the total colour D3-brane charge in this interval and $Q_{\mathrm{NS}5}$ for the total NS5-brane charge, that is, $Q_{\mathrm{NS}5}=P$. Note that even if there is a striking similarity between this expression and the expression \eqref{FHol_IIB} for the free energy of the solution discussed in section 2.1 in terms of the corresponding colour and flavour charges, now $Q_{\mathrm{NS}5}$ is infinite, due to the non-compactness of the $r$ direction. In fact, an obvious way to regularise the expression in \eqref{FHol_NATD} is to take the limits $P\rightarrow \infty$, $\epsilon\rightarrow 0$ such that $(\cot^4{\epsilon})/P$ remains finite. With this regularisation the free energy would correspond to that of ${\tilde Q}_{\mathrm{D}1}$ M5-branes with $Q_{\mathrm{D}3}$ momentum charge. We finish  this section with this suggestive relation that should be the subject of a more careful study.

\section{Conclusions}\label{conclusions}

In this paper we have provided two explicit realisations of $\mathrm{AdS}_2/\mathrm{CFT}_1$ dualities  with eight supercharges, one with $\mathrm{OSp}(4^*|4)$ supergroup and a second one with $\mathrm{OSp}(8|2)$. Both realisations emerge from controlled string theory set-ups, consisting on D0 -- NS5 -- KK or D0 -- F1 brane intersections (in type IIA language). The first solution belongs to the general class of $\mathrm{AdS}_2\times S^4\times S^2\times \Sigma_2$ solutions constructed in \cite{DHoker:2007mci}, for a Riemann surface with the topology of an annulus. We have computed the corresponding holographic central charges and shown that in the regime of validity of the type IIA description they share the same universal behaviour with the electric field found in \cite{Hartman:2008dq}. It would be interesting to provide explicit checks on the field theory side of these scalings, possibly along the lines of $\mathcal{I}$-extremization (see \cite{Benini:2015eyy,Couzens:2018wnk,Gauntlett:2018dpc,Gauntlett:2019roi}),
or using the index recently derived in \cite{Dorey:2018klg} for quantum mechanical systems with $\mathrm{OSp}(4^*|4)$ superconformal symmetry. We have shown that in the UV the non-compactness of the internal space of the $\mathrm{AdS}_2$ solutions is resolved in M-theory in terms of either M5 or M2 branes with momentum charge, thus providing 6d (2,0) or 3d $\mathcal{N}=4$ CFT completions of the respective superconformal quantum mechanics, explicitly realising deconstructed extra dimensions.

In a less controlled string theory setting, we have provided an explicit example in the class of  $\mathrm{AdS}_2\times S^4\times S^2\times \Sigma_2$ solutions in \cite{DHoker:2007mci} for which the Riemann surface is an infinite strip. This solution is related through analytic continuation to a  $\mathrm{AdS}_4\times S^2\times S^2\times \Sigma_2$ solution in the class of \cite{DHoker:2007zhm,DHoker:2007hhe}  for which the 3d dual SCFT is described by a linear quiver with gauge groups of increasing ranks, living in a D3 -- NS5 -- D5 Hanany-Witten brane set-up. Analytic continuation suggests that the superconformal quantum mechanics dual to the $\mathrm{AdS}_2\times S^4\times S^2\times \Sigma_2$ solution should arise as the IR fixed point of a supersymmetric quantum mechanics described by a linear quiver with gauge groups of increasing ranks, living in a D1 -- NS5 -- D3 brane intersection. It would be interesting to show if a one to one mapping between $\mathrm{AdS}_2\times S^4\times S^2\times \Sigma_2$ solutions and 1d SCQM similar to the one found in \cite{Assel:2011xz} between $\mathrm{AdS}_4\times S^2\times S^2\times \Sigma_2$ solutions and 3d $\mathcal{N}=4$ SCFTs  \cite{Gaiotto:2008ak} can be found. In this set-up it would be interesting to clarify whether deconstruction plays a role in the UV completion of the associated non-compact superconformal quantum mechanics. 

Finally, our interpretation for holography within AdS$_2$, suggests that a similar logic could be used to try and make sense of holography within other non-compact geometries in higher dimensions. One case are the ones generated through non-Abelian T-duality, an example of which we have encountered in this paper. Thinking along the same lines would support their holographic descriptions as emergent IR conformal phases within higher-dimensional CFTs. 

\noindent We hope to report progress in these interesting open problems in future work.

\section*{Acknowledgements}

We would like to thank Micha Berkooz and Carlos N\'u\~nez for very useful discussions. The authors are partially supported by the Spanish government grant PGC2018-096894-B-100 and by the Principado de Asturias through the grant FC-GRUPIN-IDI/2018/000174. AR is supported by CONACyT-Mexico.

\appendix
\section{Relation with $\mathrm{AdS}_4$ Backgrounds}\label{analytic}

As we have mentioned, the type IIB $\mathrm{AdS}_2$ backgrounds discussed in sections \ref{AdS2IIB} and \ref{NATD} are related to a known family of $\mathrm{AdS}_4$ type IIB solutions through an analytic continuation prescription. A first hint of this may be found at the level of the common underlying supergroup, up to a different choice of real section.
Specifically, the $\mathrm{OSp}(4^*|4)$  supergroup, besides $\mathrm{SO}(1,2)\times \mathrm{SO}(3)\times \mathrm{SO}(5)$, also admits $\mathrm{SO}(3,2)\times \mathrm{SO}(3)\times \mathrm{SO}(3)$ as a bosonic subgroup. This is the group of isometries of an $\mathrm{AdS}_4\times S^2\times S^2$ solution, dual to a 3d SCFT with 8 supercharges. These solutions were classified in \cite{DHoker:2007zhm,DHoker:2007hhe}, where they were indeed associated to the $\mathrm{OSp}(4^*|4)$  supergroup. In this section we show the details of the analytic continuation relating the solution in section \ref{AdS2IIB} with an $\mathrm{AdS}_4\times S^2\times S^2$ solution in  the class in \cite{DHoker:2007zhm,DHoker:2007hhe}. The $\mathrm{AdS}_4\times S^2\times S^2$ solution related to the solution constructed through non-Abelian T-duality in section \ref{NATD} can be worked out in a very similar way. The result is the background discussed at length in \cite{Lozano:2016wrs}, obtained from $\mathrm{AdS}_4\times S^7/(\mathbb{Z}_k\times \mathbb{Z}_{k'})$ through a chain of dualities. 
\begin{table}[ht]
	\begin{center}
		\begin{tabular}{| l | c | c | c | c| c | c| c | c| c | c |}
			\hline		    
			& 0 & 1 & 2 & 3 & 4 & 5 & 6 & 7 & 8 & 9 \\ \hline
			D3 & $\times$ & $\times$ & $\times$&  &  &  & $\times$  &   &   &   \\ \hline
			D5 & $\times$ & $\times$ &$\times$  &$\times$  & $\times$ &$\times$   &  &  &  &   \\ \hline
			NS5 & $\times$ & $\times$ & $\times$ &  &  &  &   & $\times$  &$\times$   &$\times$   \\ \hline
		\end{tabular} 
	\end{center}
	\caption{$\frac14$-BPS intersection involving D3, D5 and NS5 branes.  The intersection is $\mathrm{SO}(3)\times \mathrm{SO}(3)$ invariant due to rotational symmetry along $(x^3,x^4,x^5)$ and $(x^7,x^8,x^9)$. An $\mathrm{AdS}_4$ vacuum with the same isometries emerges in the near horizon limit.}   
	\label{D3-D5-NS5}	
\end{table} 

We focus on the type IIB $\mathrm{AdS}_2$ background analysed in section \ref{AdS2IIB}.  Acting with the following ``quadruple'' analytic continuation
\begin{equation}
ds^2_{\mathrm{AdS}_2}\rightarrow -ds^2_{{\tilde S}_2}\, , \qquad ds^2_{S^4}\rightarrow -ds^2_{\mathrm{AdS}_4}\, , \qquad \alpha\rightarrow -ir-\frac{\pi}{2}\, , \qquad e^\Phi\rightarrow i e^\Phi\, ,
\end{equation}
on the background \eqref{NH_D1D3NS5}-\eqref{D1D3NS5_horizon}, plus the change of variables $\cosh^{-1}r=\sin{\beta}$,
one finds the following $\mathrm{AdS}_4\times S^2\times S^2\times \Sigma_2$ solution
\begin{eqnarray}
&&ds_{10}^2=\ell^2\sin{\beta}\left(ds^2_{\mathrm{AdS}_4}+4d\beta^2+\cos^2{\beta}ds^2_{S^2}+\sin^2{\beta}ds^2_{{\tilde S}^2}+\frac{R_0^2}{\sin^2{\beta}\cos^2{\beta}}dy^2\right)\,, \label{metricAdS4}\\
&&e^{\Phi}=\frac{Q_{\mathrm{NS}5}}{Q_{\mathrm{D}1}}\tan{\beta}\,,\\
&&H_{(3)}=-Q_{\mathrm{NS}5}dy\wedge {\rm vol}_{\left(S^2\right)}\,,\\
&&F_{(3)}=Q_{\mathrm{D}1}{\rm vol}_{\left({\tilde S}^2\right)}\wedge dy\,,\\
&&F_{(5)}=(1+\star_{10})\left(3\pi Q_{\mathrm{D}3}{\rm vol}_{\left(\mathrm{AdS}_4\right)}\wedge dy\right)\,. \label{F5AdS4}
\end{eqnarray}
After the analytical continuation the D1-brane charge becomes D5-brane charge, and the resulting $\mathrm{AdS}_4$ vacuum is supported by D3, D5 and NS5-brane charges. This solution was found originally in \cite{Cvetic:2000cj}, as the near horizon limit of a D3 -- NS5 -- D5 brane intersection.
This brane set-up consists on D3-branes wrapped on the $y$ direction, stretched between NS5-branes, located at fixed values of $y$, with additional perpendicular D5-branes. The configuration is depicted in Table \ref{D3-D5-NS5}. 

\begin{figure}
\centering
\includegraphics[scale=0.6]{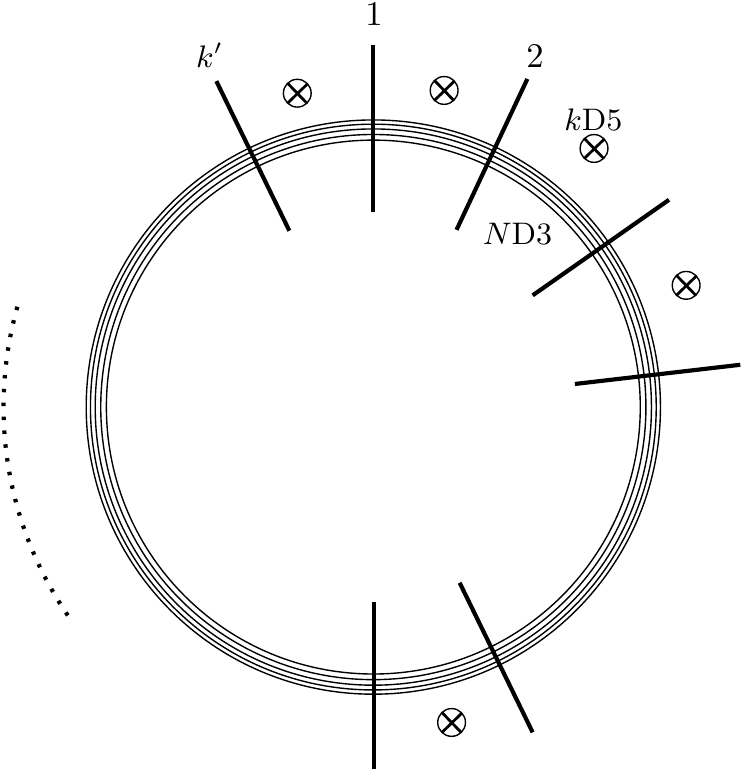}
\caption{Brane set-up underlying the $\mathrm{AdS}_4\times S^2\times S^2$ solution.}
\label{brane-set-upAdS4}
\end{figure}

The near horizon limit of the solution describing the D3 -- NS5 -- D5 brane intersection gives rise to the background given by \eqref{metricAdS4}-\eqref{F5AdS4}, which fits in the classification of $\mathrm{AdS}_4\times S^2\times S^2\times \Sigma_2$ solutions to type IIB in \cite{DHoker:2007zhm,DHoker:2007hhe}, for the choice of harmonic functions
\begin{equation}
h_1=\frac{\ell^2}{2}\frac{Q_{\mathrm{D}1}}{Q_{\mathrm{NS}5}}\cos^2{\beta}\, , \qquad h_2=\frac{\ell^2}{2}\sin^2{\beta}
\end{equation}
and
\begin{equation}
h_1^D=-\frac12 Q_{\mathrm{D}1} \, y\, , \qquad h_2^D=-\frac12 Q_{\mathrm{NS}5}\, y\, .
\end{equation}
This solution was discussed at length in \cite{Assel:2012cj} (see also Appendix A of \cite{Lozano:2016wrs}), 
as a concrete example of the one to one mapping found in \cite{Assel:2011xz} between $\mathrm{AdS}_4\times S^2\times S^2\times \Sigma_2$ solutions in the classification of \cite{DHoker:2007hhe}\footnote{Note that within the aforementioned class one should also find the analytic continuation of the $\mathrm{AdS}_4$ background found in \cite{Inverso:2016eet}, which is characterised by a linear dilaton profile obtained through an $\mathrm{SL}(2,\mathbb{Z})$ duality twist.} and 3d SCFTs with 8 supercharges \cite{Gaiotto:2008ak}. 
For this solution the Riemann surface is an annulus, with boundaries at $\beta=0,\pi/2$, where smeared D5 and NS5 branes are located. The associated quiver, depicted in Figure \ref{brane-set-upAdS4}, becomes circular.
After T-duality along the $y$-circle and uplift to eleven dimensions the solution becomes 
$\mathrm{AdS}_4\times S^7/(\mathbb{Z}_k\times \mathbb{Z}_{k'})$. Therefore, it 
describes $N$ M2-branes at a $\mathbb{C}^4/(\mathbb{Z}_k\times \mathbb{Z}_{k'})$ singularity, with $N$, $k$ and $k'$ given by the numbers of D3, D5 and NS5 branes in the type IIB set-up.  Consistently, the $\mathrm{AdS}_4\times S^7/(\mathbb{Z}_k\times \mathbb{Z}_{k'})$ background is also related to the $\mathrm{AdS}_7/(\mathbb{Z}_k\times \mathbb{Z}_{k'})\times S^4$ background obtained by uplifting the $\mathrm{AdS}_2\times S^4\times S^2$ solution in section \ref{AdS2IIB} through analytical continuation.

 \end{document}